\documentclass[10pt,twocolumn,twoside]{IEEEtran}

\usepackage{cite}
\usepackage{amsmath}
\usepackage{amssymb}
\usepackage{fontenc}
\usepackage{color}

\usepackage{algorithm}
\usepackage{algorithmic}
\usepackage{subfigure}
\allowdisplaybreaks [3]
\usepackage{graphicx}
\usepackage[caption=false,font=footnotesize]{subfig}
\usepackage{url}
\usepackage{booktabs}
\usepackage{balance}

\usepackage{xcolor}

\begin{document}

\title{Intelligent Hybrid Resource Allocation in MEC-assisted RAN Slicing Network}

\author{
        Chong~Zheng,~\IEEEmembership{Student~Member,~IEEE},
        Yongming~Huang,~\IEEEmembership{Senior Member,~IEEE},\\
        Cheng~Zhang,~\IEEEmembership{Member,~IEEE},
        and~Tony~Q.~S.~Quek,~\IEEEmembership{Fellow,~IEEE}

\thanks{Manuscript received XXX XX, 2023; revised XXX XX, XXXX; accepted XXX XX, XXXX. Date of publication XXX XX, XXXX; date of current version XXX XX, XXXX. This work was supported in part by the xxx under Grant xx. (Corresponding author: Y.~Huang.)}

\thanks{C.\ Zheng, Y.\ Huang, and C.\ Zhang are with the School of Information Science and Engineering, Southeast University, Nanjing 210096, China, and also with Purple Mountain Laboratories, Nanjing 211111, China (e-mail: \{czheng; huangym; zhangcheng\_seu\}@seu.edu.cn).}
\thanks{T.~Q.~S.~Quek is with the Information System Technology and Design Pillar, Singapore University of Technology and Design, Singapore 487372 (e-mail: tonyquek@sutd.edu.sg).}

}

\markboth{TRANSACTIONS ON VEHICULAR TECHNOLOGY,~Vol.~XX, No.~X, XXX~XXXX}%
{Zheng \MakeLowercase{\textit{et al.}}: INTELLIGENT HYBRID RESOURCE ALLOCATION IN MEC-ASSISTED RAN SLICING  NETWORK}

\maketitle
\begin{abstract}
Mobile edge computing (MEC) combining radio access network (RAN) slicing shows tremendous potential in satisfying diverse service level agreement (SLA) demands in future wireless communications. Since the limited computing capacities of MEC servers as well as the limited transmission capacities of wireless communication links, efficient hybrid resource allocation (RA) from the perspective of computing and transmission resources is crucial to maintain high SLA satisfaction rate (SSR). However, in cooperative multi-node MEC-assisted RAN slicing systems, the complexity of the multi-node cooperation in spatial dimension as well as the contextual correlation of system state in time dimension pose significant challenges to the hybrid RA policy optimization. In this paper, we aim to maximize the SSR for heterogeneous service demands in the cooperative MEC-assisted RAN slicing system by jointly considering the multi-node computing resources cooperation and allocation, the transmission resource blocks (RBs) allocation, and the time-varying dynamicity of the system. To this end, we abstract the system into a weighted undirected topology graph and, then propose a recurrent graph reinforcement learning (RGRL) algorithm to intelligently learn the optimal hybrid RA policy. Therein, the graph neural network (GCN) and the deep deterministic policy gradient (DDPG) is combined to effectively extract spatial features from the equivalent topology graph. Furthermore, a novel time recurrent reinforcement learning framework is designed in the proposed RGRL algorithm by incorporating the action output of the policy network at the previous moment into the state input of the policy network at the subsequent moment, so as to cope with the time-varying and contextual network environment. In addition, we explore two use case scenarios to discuss the universal superiority of the proposed RGRL algorithm. Simulation results demonstrate the superiority of the proposed algorithm in terms of the average SSR, the performance stability, and the network complexity.
\end{abstract}


\bigskip

\section{INTRODUCTION}
\label{sec1}

\IEEEPARstart{A}{long} with the thriving development of the fifth generation (5G) and beyond communication technologies, people's life experience has achieved a further leap while various emerging service requirements have been soaring at an unprecedented pace. To satisfy the diverse service level agreement (SLA) demands, 5G and beyond network are supposed to support a variety of application scenarios \cite{WangZ22,Foukas17,ZhouX16,YouX21}. According to international telecommunication union-radio communication sector (ITU-R) recommendations, three typical service scenarios, i.e., enhanced mobile broadband (eMBB), massive machine type communication (mMTC), and ultra-reliable and low-latency communication (uRLLC), should be involved \cite{ITU-R17,ZaidiAA18,KovalchukovR22,AzariA19}. A promising technology for the heterogeneous service demands can be found by mobile edge computing (MEC)-assisted radio access network (RAN) slicing \cite{TalebT17,ChienHT19}. On the one hand, network slicing (NS) abstracts and virtualizes the physical common multi-domain infrastructure into various network slices so as to realize the logical isolation of physical resources and ensure service customization \cite{WangZ22}. On the other hand, MEC framework provides abundant computing and caching resources within network edge by equipping the edge nodes with processing servers, thereby supporting edge services with specific requirements as well as significantly improving the service quality \cite{FengJ20,ZhengC22}. 

Although the MEC-assisted RAN slicing provides supports for diverse SLA demands, efficiently and intelligently utilization of the hybrid resources to maintain high SLA satisfaction rate (SSR) is still necessary and challenging. The explorations of hybrid resources allocation (RA) in MEC-assisted RAN slicing have been established in many works. The authors in \cite{Opti19,Combine22} investigate the hybrid RA in the scenario of NS merging single MEC node from the perspectives of queue optimization and machine learning, respectively. The advantages of the hybrid resources allocation optimization in SSR improvement is proved in \cite{Opti19,Combine22}, while the network architecture with only single MEC node is too simple to fully exploit potentialities of the MEC-assisted RAN slicing system. Consequently, the NS with multiple MEC nodes is considered in \cite{Dyna20,Utility22,Blockchain23}. For example, the authors in \cite{Utility22} investigate the joint optimization of transmission, computing and caching resources in multi-access edge network slicing and propose a deep reinforcement learning (DRL)-based approach to maximize the utility of the system while ensuring the quality of service (QoS). Nevertheless, the cooperation, especially in computing cooperation, between MEC nodes can further improve the resource utilization efficiency of the MEC-assisted RAN slicing system, which has been exploited in \cite{Delay21} but not studied in \cite{Dyna20,Utility22,Blockchain23}. In \cite{Delay21}, the weighted sum gap between the observed latency and the tolerant latency is minimized by jointly optimizing computing and transmission RA in the collaborative multi-node MEC-assisted RAN slicing network. The authors in \cite{Delay21} leverage on the traditional convex optimization methods, i.e., fractional programming and augmented lagrangian, to optimize the instantaneous RA and SSR. However, when the network scale increases, computing cooperations of multiple nodes in space dimension will bring many challenges such as extremely high complexity and NP-hard problems to the RA policy optimization based on traditional optimization methods.

In fact, multi-node computing collaborations make the network exhibit topological graph structure characteristics in the spatial dimension, while graph learning, as an attractive option for processing graph data \cite{WuZ21}, becomes a powerful method to promote the RA policy optimization by effectively extracting spatial features from the network topology \cite{WangZ22,EisenM20,WangD23}. In \cite{WangZ22}, authors consider a generalized optimal RA problem in a multi-access wireless network and develop an aggregation graph neural network (Agg-GNN) to aggregate spatial state information for the potentially asynchronous graph of the wireless network. 
By further combining the GNN with the convolutional neural network whose dimensionality is small and does not scale with network size, authors in \cite{EisenM20} propose a random edge graph convolutional network (GCN) to learn the optimal RA policy in a multi-node RAN. Similarly, authors in \cite{WangD23} formulate the joint optimization of hybrid resources in the MEC-enabled Internet of Things (IoT) as a multiagent decision problem and then propose a deep graph convolution reinforcement learning method to learn optimal RA policy. These works demonstrate the advantages of GCNs with spatial feature extraction capabilities in improving the performance of RA policy optimization in collaborative multi-node networks. 

On the other hand, dynamic environments of edge networks, e.g., the dynamic generation of user tasks, the dynamic resource capabilities, and etc., cause the system state to be contextual in time dimension. The authors in \cite{Design20} are aware of the importance of the time-dimensional context and model a Markov process in the collaborative multi-node MEC-assisted RAN slicing system so as to capture the time-varying features of the system. Nevertheless, authors in \cite{Design20} mainly focus on the computing RA and ignore the the hybrid RA including the communication resources. On the contrary, many other works, e.g., \cite{HuaY20,YanM19,CuiY22}, consider the influence of time dimension and study the single transmission RA without considerations about computing RA. Regrettably, all these works \cite{Design20,HuaY20,YanM19,CuiY22} only consider a single RA of computing or transmission resources. Commonly in \cite{Design20,HuaY20,YanM19,CuiY22}, the DRL is adopted as an effective method to handle the time-contextual system state. On this basis, some works, e.g., \cite{Opti19,Utility22,HaoM21,GongY23}, have further explored advantages of the DRL in optimizing the hybrid RA policy in the MEC-assisted RAN slicing system and maximize the long term system performance under considerations of dynamic environments. However, these related works simply exploited the Markov chain property in DRL to capture the time-dimensional context of the system state without further considerations of the structure design of DRL for feature extractions in time dimension. In \cite{LiR20}, authors design a recurrent structure by embedding a classical long short-term memory (LSTM) into the DRL to improve the ability of agent in capturing temporal features. However, the LSTM is a recurrent structure designed for the natural language processing. When applied into the communication problem, the adaptability to communication problems of the LSTM is limited and moreover the the complexity of LSTM is extremely high. In this paper, we design a novel time-recurrent structure for the DRL to extract temporal features more effectively in the investigated communication problem.


As a matter of fact, when both the space dimension and time dimension are involved, the hybrid RA policy for the collaborative multi-node MEC-assisted RAN slicing system is required not only to perform effective RA actions on different nodes in space, but also to consider the contextual relationship in time dimension so as to ensure the system a high SSR over a continuous period of time. Consequently, these stringent requirements pose tough challenges to the policy optimization of the hybrid RA. In this paper, we investigate the hybrid RA optimization in the collaborative multi-node MEC-assisted RAN slicing system by jointly considering the computing cooperations among multiple MEC nodes and the transmission resource block (RB) allocation as well as the time-varying dynamic environment of the system. We strive to maximize the SSR for heterogeneous service demands in the considered communication system and propose a recurrent graph reinforcement learning (RGRL) algorithm to intelligently learn the optimal hybrid RA policy. 

Specifically, the contributions are summarized as follows:
\begin{itemize}
	\item The hybrid computing and communication resources allocation problem for heterogeneous service scenarios in the cooperative multi-node MEC-assisted RAN slicing system is studied. By jointly considering computing cooperations of multiple MEC nodes in space dimension and contextual relevances of the dynamic network environment in time dimension, we design the hybrid allocation policy to maximize the SSR of heterogeneous service requirements in the system.
	
	\item We abstract the multi-node slicing system into a weighted undirected topology and then propose an RGRL algorithm to intelligently optimize the hybrid allocation policy. Therein, we combine the GCN with the deep deterministic policy gradient (DDPG) to address the policy optimization challenges caused by collaborations of the heterogeneous large-scale network in space dimension. In addition, a novel time recurrent reinforcement learning framework is designed in the proposed RGRL algorithm by incorporating the action output of the policy network at the previous moment into the state input of the policy network at the subsequent moment, so as to cope with the time-varying and contextual network environment.
	
	\item We further investigate two use cases, i.e., single-node MEC-aided RAN slicing scenario and non-cooperative multi-node MEC-aided RAN slicing scenario, to explore the universality of the proposed RGRL algorithm. Numerical simulations in the main research scenario as well as two use cases demonstrate the universal superiority of the RGRL algorithm in SSR performance.
\end{itemize}

The rest of this paper is orgnized as follows. Section \ref{sec_2} describes the system model. Section \ref{sec_3} introduces the problem formulation and methodology. Then, two use cases are discussed and analyzed in Section \ref{sec_4}. In Section \ref{sec_5}, simulation results are discussed. Finally, conclusions are drawn in Section \ref{sec_6}.

\begin{figure*}[!tpbh]
\centering
\includegraphics[width=0.35\textwidth, angle=90]{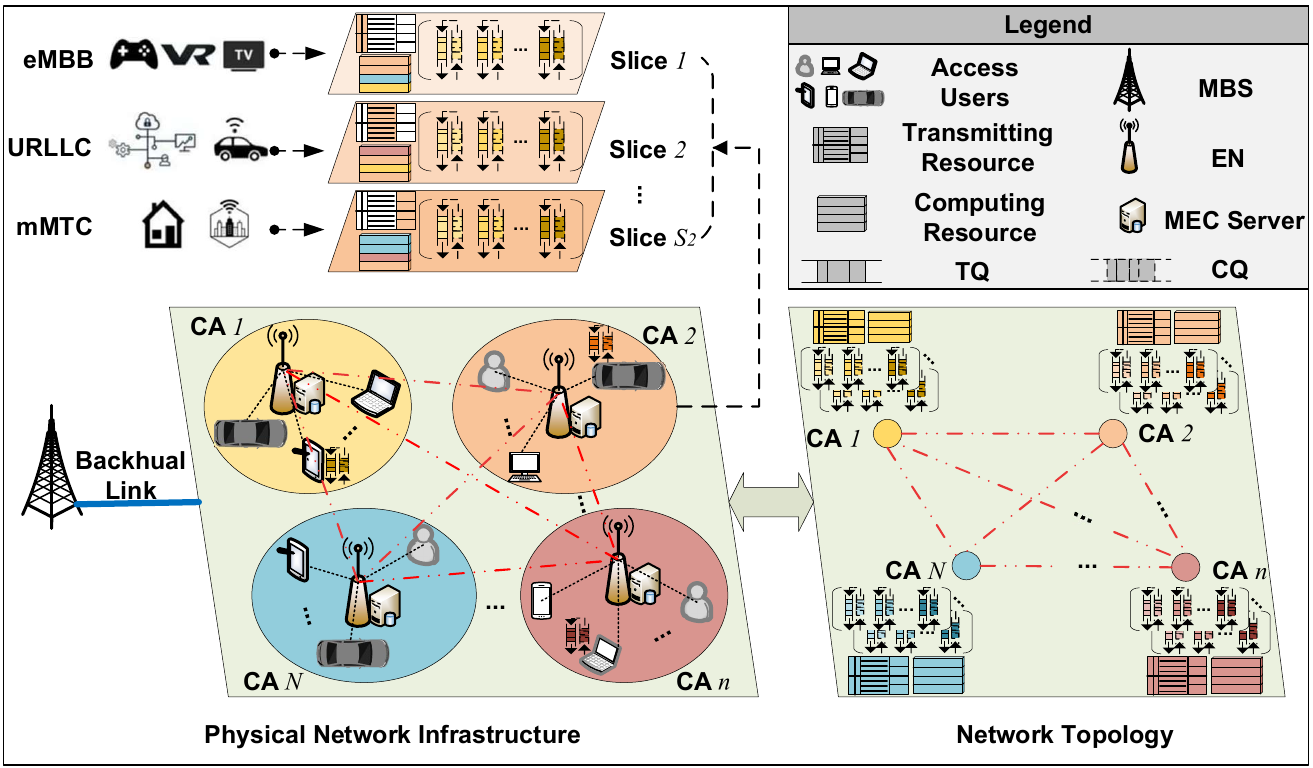}
\caption{System model.}
\label{sys}
\end{figure*}

\section{SYSTEM MODEL}
\label{sec_2}
Fig. \ref{sys} illustrates a collaborative multi-node MEC-assisted RAN slicing scenario with one macro base station (MBS), $N$ edge nodes (ENs), and $U$ access users (AUs), where each EN is equipped with a MEC server and can cooperate with neighboring nodes in task computing. Assume that each EN is randomly distributed in a rectangular region of $d_0 \times d_0$. Consideing the computing cooperations and cooperation costs among ENs, the network topology of the system can be expressed as a weighted undirected graph. Let $A_{\rm max}$ denote the maximum number of neighbors for each EN. We have $1\leq A_{{\rm max}}\leq N$. We assume that EN $n$ can divide $S_n$ NSs to provide different communication services, i.e., eMBB, uRLLC, and mMTC, for users in the coverage area (CA) $n$. The CA of each EN is assumed as a circle with radius of $d_r$, and users in each CA are randomly distributed in a ring area with radius of $\left[d_{min},d_{max}\right]$ and center of the EN, where $0\leq d_{min}<d_{max}\leq d_{r}$. It is reasonable to assume that each EN and its neighboring nodes can cooperate in task computing, but the results transmission can only be completed by each EN itself.


\subsection{Task Generation Model}\label{Ta_ge}

Let $\mathcal{U}^{n}=\left\{ \left.\mathcal{U}_{s}^{n}\right|s=1,2,\cdots,S_{n}\right\}$ denote the user set of the EN $n$, where $\mathcal{U}_{s}^{n}$ represents the user set under the $s$-th slice of the EN $n$. At the beginning of each time slot $t$, each user $u\in\mathcal{U}_{s}^{n}$ generates a computing task request at an arrival rate $\kappa_{u}^{n,s}$,  where $0<\kappa_{u}^{n,s}\leq1$. Suppose that $q_{u}^{n,s}\left(t\right)$ is a binary variable that is 1 if user $u$ generates a task request at time $t$ and 0 otherwise. When $q_{u}^{n,s}\left(t\right)=1$, the size of the requested computing task at time $t$, i.e., $d_{u}^{n,s}\left(t\right)$, follows a certain mathematical distribution. As adopted in many related works \cite{Distributed22,Hybrid21}, we assume that $d_{u}^{n,s}\left(t\right)$ obeys a Pareto distribution with the shape parameter $\zeta_{u}^{n,s}$ and the threshold parameter $\delta_{u}^{n,s}$. Hence, the probability density distribution of $d_{u}^{n,s}\left(t\right)$ can be expressed as
\begin{equation} \label{e1}
p\left(d_{u}^{n,s}\left(t\right)\right)=\begin{cases}
0, & d_{u}^{n,s}\left(t\right)\leq\delta_{u}^{n,s}\\
\frac{\zeta_{u}^{n,s}\cdot\left(\delta_{u}^{n,s}\right)^{\zeta_{u}^{n,s}}}{\left(d_{u}^{n,s}\left(t\right)\right)^{\zeta_{u}^{n,s}+1}}, & d_{u}^{n,s}\left(t\right)>\delta_{u}^{n,s}
\end{cases}.
\end{equation}

\subsection{Hybrid Resource Allocation Model}

Let $C_{\rm B}$ and $Z_{\rm B}$ respectively denote the computing frequency and transmission RBs of each EN.  Let $\mathbf{c}^{n}\left(t\right)=\left[c_{1}^{n}\left(t\right),\cdots,c_{N}^{n}\left(t\right)\right]$ denote the computing resource allocation of the EN $n$ for all the ENs at time $t$, where $c_{j}^{n}\left(t\right),j=1,\cdots,N$ represents the proportion of computing resources allocated by EN $n$ to EN $j$ and $\sum_{j=1}^{N}c_{j}^{n}\left(t\right)=1$. Considering the constraint of the maximum number of neighbor nodes, the actual proportion of computing resources allocated by EN $n$ to EN $j$ at time $t$ can be renewed as 
\begin{equation} \label{e2}
\overline{c}_{j}^{n}\left(t\right)=\frac{\lambda_{j}^{n}c_{j}^{n}\left(t\right)}{\sum_{j=1}^{N}\lambda_{j}^{n}c_{j}^{n}\left(t\right)}.
\end{equation}
$\lambda_{j}^{n}$ is the element of $\mathbf{A}=\left[\lambda_{j}^{i}\right]_{i,j=1}^{N}$ which is the adjacency matrix with self-connections of the network topology. Moreover, the computing resources allocated to EN $n$ at time $t$ can be expressed as
\begin{equation} \label{e3}
C^{n}\left(t\right)=\sum_{j=1}^{N}\delta_{j}^{n}\overline{c}_{j}^{n}\left(t\right)C_{{\rm B}}.
\end{equation}
$0\leq\delta_{j}^{n}=\delta_{n}^{j}\leq1$ is the weight of the edge between EN $n$ and EN $j$, which is used to measure the cost of cooperation between these two nodes and can be defined by
\begin{equation}
\delta_{j}^{n}=\begin{cases}
\frac{d_{{\rm min}}}{d_{j}^{n}}\vartheta, & d_{j}^{n}\neq0\\
1, & d_{j}^{n}=0
\end{cases},
\end{equation}
where $\vartheta$ is the penalty factor for computing cooperation between ENs, and $d_{j}^{n}$ is the euclidean distance between EN $n$ and EN $j$. Especially, $d_{j}^{n}=0$ when $n=j$ and $d_{j}^{n}=+\infty$ when $\lambda_{j}^{n}=0$. $d_{{\rm min}}$ is the shortest edge length among all edges in the topological graph. We can obtain the weighted adjacency matrix with self-connections, denoted as $\mathbf{\widetilde{A}}=\left[\lambda_{j}^{i}\cdot\delta_{j}^{i}\right]_{i,j=1}^{N}$. Then, the computing resources allocated to the $s$-th slice of EN $n$ at time $t$ can be given by 
\begin{equation} \label{e4}
C^{n,s}\left(t\right)=c^{n,s}\left(t\right)C^{n}\left(t\right),
\end{equation}
where $c^{n,s}\left(t\right)$ represents the allocation proportion of EN $n$ to its $s$-th slice at time $t$ and $\sum_{s=1}^{S_{n}}c^{n,s}\left(t\right)=1$. Finally, the computing resources allocated to each NS user $u\in\mathcal{U}_{s}^{n}$ at time $t$ can be expressed as
\begin{equation} \label{e5}
C^{n,s,u}\left(t\right)=c^{n,s,u}\left(t\right)C^{n,s}\left(t\right),
\end{equation}
where $c^{n,s,u}\left(t\right)$ is the allocation proportion and $\sum_{u\in\mathcal{U}_{s}^{n}}c^{n,s,u}\left(t\right)=1$. Furthermore, according to \cite{Hybrid21}, the computing speed obtained by NS user $u$ at time $t$ can be given by
 \begin{equation}
RC_{u}^{n,s}\left(t\right)=\frac{\varpi}{C^{n,s,u}\left(t\right)},
\end{equation}
where $\varpi$ is the central processing unit (CPU) cycles involved in computing one-bit data.

After the computing, the calculation results need to be transmitted to the corresponding NS users, which involves allocations of transmission RBs. Let $\mathbf{z}^{n}\left(t\right)=\left[z^{n,1}\left(t\right),\cdots,z^{n,S_{n}}\left(t\right)\right]$ denote the transmission RBs allocation of the EN $n$ for its NSs at time $t$, where $z^{n,s}\left(t\right)$ denote the proportion of transmission RBs allocated by EN $n$ to its $s$-th slice at time $t$, and $\sum_{s=1}^{S_{n}}z^{n,s}\left(t\right)=1$. Then we can obtain the following transmission RBs allocated to the $s$-th slice of EN $n$ at time $t$
\begin{equation} \label{e6}
Z^{n,s}\left(t\right)=z^{n,s}\left(t\right)Z_{{\rm B}}.
\end{equation}
Moreover, let $z^{n,s,u}\left(t\right)$ denote the transmission RBs allocation proportion of each slice $s$ under EN $n$ to its NS user $u$ at time $t$. Hence, the the transmission RBs obtained by each NS user $u\in\mathcal{U}_{s}^{n}$ at time $t$ can be expressed as
\begin{equation} \label{e7}
Z^{n,s,u}\left(t\right)=z^{n,s,u}\left(t\right)Z^{n,s}\left(t\right),
\end{equation}
where $\sum_{u\in\mathcal{U}_{s}^{n}}z^{n,s,u}\left(t\right)=1$. Subsequently, we can have the data transmission rate obtained by user $u$ at time $t$.
\begin{equation} \label{e8}
RT_{u}^{n,s}\left(t\right)=Z^{n,s,u}\left(t\right)W_{{\rm B}}\textrm{log}_{2}\left(1+\gamma_{u}^{n,s}\left(t\right)\right).
\end{equation}
$W_{{\rm B}}$ represents the bandwidth of each RB and $\gamma_{u}^{n,s}\left(t\right)$ is the signal-to-noise ratio (SNR) between $u$ and the $s$-th slice of EN $n$ at time $t$.
\begin{equation}
\gamma_{u}^{n,s}\left(t\right)=\frac{P_{u}^{n,s}\left(t\right)h_{u}^{n,s}\left(t\right)}{N_{0}Z^{n,s,u}\left(t\right)W_{{\rm B}}}.
\end{equation}
where $N_0$ denotes the noise power spectral density (NPSD). $h_{u}^{n,s}\left(t\right)=g_{u}^{n,s}\left(t\right)\cdot\left(d_{u}^{n}\right)^{-\beta}$ presents the average channel gain between user $u$ and the $s$-th slice of EN $n$ at time $t$, where $g_{u}^{n,s}\left(t\right)$ is the Rayleigh fading parameter, $d_{u}^{n,s}$ denotes the distance between the user and the EN, and $\beta$ is the path loss exponent. Moreover, $P_{u}^{n,s}\left(t\right)=P_{{\rm B}}Z^{n,s,u}\left(t\right)$ is the transmission power, where $P_{{\rm B}}$ is the transmission power of each RB.

\subsection{Queue Model}
The computing requests are continuously generated by NS users and sequentially uploaded to the corresponding NS for calculation. Therefore, a task request queue and a slice-level computing queue (CQ) set will be formed at each NS user and each NS, respectively. Let $\mathcal{Q}_{s}^{n}\left(t\right)=\cup_{u\in\mathcal{U}_{s}^{n}}\mathcal{Q}_{u}^{n,s}\left(t\right)$ represent the slice-level CQ set at the $s$-th NS of EN $n$ at time $t$, where $\mathcal{Q}_{u}^{n,s}\left(t\right)=\left\{ \left.d_{u}^{n,s}\left(\tau\right)\right|\tau=t-T,\cdots,t\right\}$ is the CQ of user $u\in\mathcal{U}_{s}^{n}$ at time $t$. Suppose $T$ is a time window of observation. Furthermore, we can have the node-level CQ set at EN $n$, i.e., $\mathcal{Q}_{n}\left(t\right)=\cup_{s=1}^{S_{n}}\mathcal{Q}_{s}^{n}\left(t\right)$. Let $Q_{u}^{n,s}\left(t\right)$ denote the size of $\mathcal{Q}_{u}^{n,s}\left(t\right)$. We can have
\begin{equation}
Q_{u}^{n,s}\left(t\right)={\rm max}\left\{ Q_{u}^{n,s}\left(t-1\right)+d_{u}^{n,s}\left(t\right)-RC_{u}^{n,s}\left(t\right)\nabla t,0\right\},
\end{equation}
where $\nabla t$ is the duration of the time slot $t$. There is also a transmission queue (TQ) set at each NS to transmit the calculation results to corresponding NS users through radio links with various number of RBs. The slice-level TQ set at each NS is denoted as $\mathcal{V}_{s}^{n}\left(t\right)=\cup_{u\in\mathcal{U}_{s}^{n}}\mathcal{V}_{u}^{n,s}\left(t\right)$. Similarly, there will be a node-level TQ set at each EN and a slice-level TQ set at each NS, respectively denoted as $\mathcal{V}_{n}\left(t\right)=\cup_{s=1}^{S_{n}}\mathcal{V}_{s}^{n}\left(t\right)$ and $\mathcal{V}_{s}^{n}\left(t\right)=\cup_{u\in\mathcal{U}_{s}^{n}}\mathcal{V}_{u}^{n,s}\left(t\right)$. $\mathcal{V}_{u}^{n,s}\left(t\right)=\left\{ \left.v_{u}^{n,s}\left(\tau\right)\right|\tau=t-T,\cdots,t\right\}$ is the TQ for user $u\in\mathcal{U}_{s}^{n}$ at time $t$, where $v_{u}^{n,s}\left(\tau\right)$ is the output size of the CQ at time $\tau$. For facilitating the problem, we assume the input and output sizes of one computing task are equal. Then we define the size of $\mathcal{V}_{u}^{n,s}\left(t\right)$ as $V_{u}^{n,s}\left(t\right)$.
\begin{equation}
\begin{array}{l}
V_{u}^{n,s}\left(t\right)\\={\rm max}\left\{ V_{u}^{n,s}\left(t-1\right)+\left(RT_{u}^{n,s}\left(t\right)-RC_{u}^{n,s}\left(t\right)\right)\nabla t,0\right\}.
\end{array}
\end{equation}

\subsection{Performance Metric Model}

Under the hybrid resource allocation policy of the system at time $t$, the latency for task $d_{u}^{n,s}\left(t\right)$ to be completed, denoted as $l_{u}^{n,s}\left(t\right)$, can be given by
\begin{equation}
l_{u}^{n,s}\left(t\right)=\frac{Q_{u}^{n,s}\left(t\right)+V_{u}^{n,s}\left(t\right)}{{\rm min}\left(RC_{u}^{n,s}\left(t\right),RT_{u}^{n,s}\left(t\right)\right)}.
\end{equation}
Service level agreement (SLA) demands on latency is used for the metric. Let $M_{u}^{n,s}\left(t\right)$ indicates wether the service latency of task $d_{u}^{n,s}\left(t\right)$ meets the SLA requirement or not.
\begin{equation}
M_{u}^{n,s}\left(t\right)=\begin{cases}
1, & l_{u}^{n,s}\left(t\right)\leq l_{{\rm req}}^{n,s}{\rm and}q_{u}^{n,s}\left(t\right)=1\\
0, & {\rm otherwise}
\end{cases}
\end{equation}
where $l_{{\rm req}}^{n,s}$ is the SLA demand on latency for the slce $s$ of EN $n$.
Noted that, diffirential $l_{{\rm req}}^{n,s}$ are emerged to satisfy the different types of service scenarios contained in the investigated network slicing system. Finally, the system performance can be measured by the average SLA satisfaction rate (SSR) $U_{{\rm avg}}\left(t\right)$, denoted as
\begin{equation}\label{e15}
U_{{\rm avg}}\left(t\right)=\frac{\sum_{n=1}^{N}\sum_{s=1}^{S_{n}}\sum_{u\in\mathcal{U}_{s}^{n}}M_{u}^{n,s}\left(t\right)}{\sum_{n=1}^{N}\sum_{s=1}^{S_{n}}\sum_{u\in\mathcal{U}_{s}^{n}}q_{u}^{n,s}\left(t\right)}.
\end{equation}

\begin{figure*}[!t]
\centering
\includegraphics[width=0.4\textwidth, angle=90]{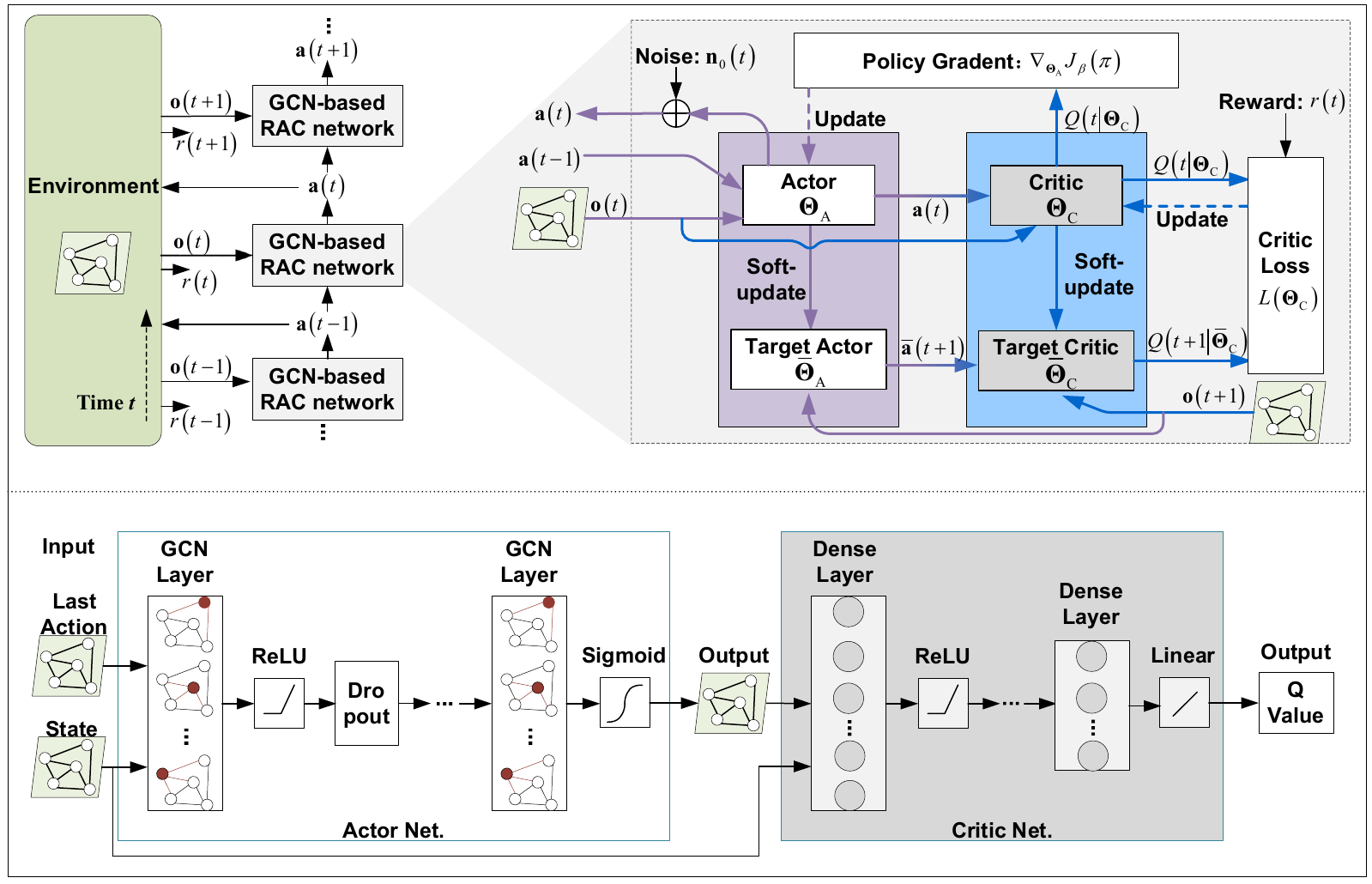}
\caption{Framework of the proposed RGRL algorithm.}
\label{RGC-DDPG}
\end{figure*}

\section{PROBLEM FORMULATION AND METHODOLOGY}\label{sec_3}
\subsection{Problem Formulation}

We maximize the long-term average SSR over a continuous period of time by optimizing the hybrid resources allocation policy.
\begin{subequations}\label{Pro_form}
    \begin{align}
	\quad
    \underset{\mathcal{A}}{{\rm max}} \quad
    &\underset{T\rightarrow\infty}{{\rm lim}}\sum_{\tau=0}^{T}\mathbb{E}\left[\chi^{\tau}U_{{\rm avg}}\left(t+\tau\right)\right],\label{eP_a}\\
    {\rm{s.t.}}\quad \quad
    & \sum_{j=1}^{N}c_{j}^{n}\left(t\right)=1,\forall n\in\mathcal{N}, \label{eP_b}\\
    &\sum_{s=1}^{S_{n}}c^{n,s}\left(t\right)=1,\sum_{s=1}^{S_{n}}z^{n,s}\left(t\right)=1,\forall n\in\mathcal{N},\label{eP_c}\\
    &\sum_{u\in\mathcal{U}_{s}^{n}}c^{n,s,u}\left(t\right)=1,\sum_{u\in\mathcal{U}_{s}^{n}}z^{n,s,u}\left(t\right)=1,\nonumber \\
    &\forall s\in\mathcal{S}_{n},\forall n\in\mathcal{N},\label{eP_d}
    \end{align}
\end{subequations}
where $\mathcal{N}=\left\{ 1,\cdots,N\right\}$ and $\mathcal{S}_{n}=\left\{ 1,\cdots,S_{n}\right\}$. $\mathcal{A}=\left\{ \left.\mathbf{a}\left(t\right)=\left[\mathbf{c}\left(t\right),\mathbf{z}\left(t\right)\right]\right|t=0,1,2,\cdots\right\}$ reprents the hybrid resources allocation policy. Therein, $\mathbf{c}\left(t\right)=\left[\mathbf{c}^{n}\left(t\right),\mathbf{c}^{n,s}\left(t\right),\mathbf{c}^{n,s,u}\left(t\right)\right]_{u\in\mathcal{U}_{s}^{n},s\in\mathcal{S}_{n},n\in\mathcal{N}}$ is the stack of the computing resource allocations at time $t$, where $\mathbf{c}^{n}\left(t\right)=\left[c_{j}^{n}\left(t\right)\right]_{j\in\mathcal{N}}$, $\mathbf{c}^{n,s}\left(t\right)=\left[c^{n,s}\left(t\right)\right]_{s\in\mathcal{S}_{n}}$ and $\mathbf{c}^{n,s,u}\left(t\right)=\left[c^{n,s,u}\left(t\right)\right]_{u\in\mathcal{U}_{s}^{n}}$. Similarly, $\mathbf{z}\left(t\right)=\left[\mathbf{z}^{n,s}\left(t\right),\mathbf{z}^{n,s,u}\left(t\right)\right]_{u\in\mathcal{U}_{s}^{n},s\in\mathcal{S}_{n},n\in\mathcal{N}}$ is the stack of the transmission resource allocations at time $t$, where $\mathbf{z}^{n,s}\left(t\right)=\left[z^{n,s}\left(t\right)\right]_{s\in\mathcal{S}_{n}}$ and $\mathbf{z}^{n,s,u}\left(t\right)=\left[z^{n,s,u}\left(t\right)\right]_{u\in\mathcal{U}_{s}^{n}}$. Furthermore, $\chi\in\left[0,1\right]$ is the discount factor, and the expectation is taken with respect to the measure included by the decision variables as well as the system state. 

We can observe some facts from the problem \ref{Pro_form} that the objective function is accumulated over time rather than an instantaneous function, and the solution of the problem \ref{Pro_form} is a dynamic policy over time rather than a transient one. Therefore, these facts pose significant challenges to traditional optimization methods for solving this problem \ref{Pro_form}. In this work, we exploit the deep reinforcement learning to solve the problem \eqref{Pro_form}. To this end, we convert the underlying optimization problem into a Markove decision process (MDP) which contains following four components.

\subsubsection{State}
At time $t$, the node states of each EN $n$ consists of the CQ, TQ, and channel of all users in its service region, denoted as $\mathbf{o}^{n}\left(t\right)=\left[{\rm vec}\left\{ \mathcal{Q}_{u}^{n,s}\left(t\right)\right\} ,{\rm vec}\left\{ \mathcal{V}_{u}^{n,s}\left(t\right)\right\} ,h_{u}^{n,s}\left(t\right)\right]_{u\in\mathcal{U}_{s}^{n},s\in\mathcal{S}_{n}}$, where ${\rm vec}\left\{ \cdot\right\} $ represents the vectorized mapping. All the node states as well as the weighted adjacency matrix contribute to the system state, defined as $\mathbf{o}\left(t\right)=\left[\mathbf{o}^{1}\left(t\right),\cdots,\mathbf{o}^{N}\left(t\right),\tilde{\mathbf{A}}\right]$.

\subsubsection{Action}
From above formulated problem, we already have the actions at time $t$, i.e., $\mathbf{a}\left(t\right)=\left[\mathbf{c}\left(t\right),\mathbf{z}\left(t\right)\right]$.

\subsubsection{State Transition}
Due to the existence of $h_{u}^{n,s}\left(t\right)$ in the system state, the transition probability of the system state is unavailable.

\subsubsection{Reward}
According to the optimization objective in our problem \eqref{Pro_form}, the instantaneous reward at time $t$ and the cumulative reward starting from time $t$ can be given by $r\left(t\right)=U_{{\rm avg}}\left(t\right)$ and $R\left(t\right)=\sum_{\tau=0}^{T}\chi^{\tau}r\left(t+\tau\right)$, respectively.


\subsection{Methodology}\label{method}
In this subsection, we develop a RGRL algorithm based on the graph convolution network (GCN) \cite{JiangJ20,YouY20} and the deep deterministic policy gradient (DDPG) \cite{Silver2014Deterministic, Lillicrap2015Continuous} to learn the optimal hybrid policy for computing and transmission resources allocation. The framework of the designed algorithm is illustrated in Fig. \ref{RGC-DDPG}. The entire communication system can be represented as a topological graph as the input of the algorithm network, where $\mathbf{o}^{n}\left(t\right),n\in\mathcal{N}$ and $\mathbf{\widetilde{A}}$ can be viewed as the node features and edge features of the graph, respectively.

Different from traditional reinforcement learning, the input of our proposed algorithm network, i.e., the GCN-based recurrent actor-critic (RAC) network, includes not only the current state $\mathbf{o}\left(t\right)$, but also the output of the algorithm network in the previous adjacent time slot, i.e., $\mathbf{a}\left(t-1\right)$. Specifically, at time $t$, $\mathbf{o}\left(t\right)$ and $\mathbf{a}\left(t-1\right)$ are together fed into the actor network which is a neural network containing several graph convolution layer and can be equally denoted as a parameterized actor function $\mathbf{a}\left(t\right)=\pi^{\mathbf{\Theta}_{{\rm A}}}\left(\mathbf{o}\left(t\right),\mathbf{a}\left(t-1\right)\right)$. Therein, $\pi$ is the policy on parameters $\mathbf{\Theta}_{{\rm A}}$. In addition, the propagation of a GCN layer can be defined by
\begin{equation}
f\left(\mathbf{H}^{\left(l_{{\rm a}}\right)},\mathbf{A}\right)=\varphi\left(\mathbf{D}^{-\frac{1}{2}}\mathbf{A}\mathbf{D}^{-\frac{1}{2}}\mathbf{H}^{\left(l_{{\rm a}}\right)}\mathbf{W}^{\left(l_{{\rm a}}\right)}\right).
\end{equation}
$\mathbf{D}$ is the diagonal matrix with elements defined as $\mathbf{D}_{ii}=\sum_{j}\lambda_{j}^{i}$. $\mathbf{H}^{\left(l_{{\rm a}}\right)}$ and $\mathbf{W}^{\left(l_{{\rm a}}\right)}$ represent the feature matrix and the weights, respectively, at the $l_{\rm a}$-th layer of the actor network. $\varphi$ is the activation function. Furthermore, the exploration-exploitation balance by adding noise on policy output is adopted. Thus the policy output can be renewed as
\begin{equation}\label{e18}
\mathbf{a}\left(t\right)=\pi^{\mathbf{\Theta}_{{\rm A}}}\left(\mathbf{o}\left(t\right),\mathbf{a}\left(t-1\right)\right)+\mathbf{n}_{0}\left(t\right),
\end{equation}
where $\mathbf{n}_{0}\left(t\right)$ denotes the additive white Gaussian noise with $\mathcal{CN}\sim\left(0,\sigma^{2}\right)$. Then $\mathbf{a}\left(t\right)$ acts on the system environment to change the state to $\mathbf{o}\left(t+1\right)$ while the environment feed back the $r\left(t\right)$. Subsequently, $\mathbf{o}\left(t+1\right)$ and $\mathbf{a}_{0}\left(t\right)$ will be fed to the actor network at time $t+1$ as another loop. 

Moreover, regarding to train the actor network, $\mathbf{a}\left(t\right)$ together with $\mathbf{o}^{n}\left(t\right)$ will be further send to the critic network which is also a neural network with several dense layers. Consequently, the critic outputs the Q value $Q\left(\left.t\right|\mathbf{\Theta}_{{\rm C}}\right)$ which is an estimate of the real Q value defined as
\begin{equation}
Q\left(\left.t\right|\mathbf{\Theta}_{{\rm C}}\right)=\mathbb{E}_{\pi}\left[\left.\sum_{\tau=0}^{+\infty}\chi^{\tau}U_{{\rm avg}}\left(t+\tau\right)\right|\mathbf{o}\left(t\right),\mathbf{a}\left(t\right)\right].
\end{equation}
$\mathbf{\Theta}_{{\rm C}}$ represents the parameters of the critic network. Subsequently, $Q\left(\left.t\right|\mathbf{\Theta}_{{\rm C}}\right)$ will be fed back to guide the training of the actor by calculating the policy gradient $\nabla_{\mathbf{\Theta}_{{\rm A}}}J_{\beta}\left(\pi\right)$. To better train the actor network, it is necessary to train the critic network to obtain a more accurate Q value. Specifically, we first named $\pi^{\mathbf{\Theta}_{{\rm A}}}\left(\cdot\right)$ and $Q\left(\left.\cdot\right|\mathbf{\Theta}_{{\rm C}}\right)$ online networks and, then set two target networks, i.e., $\pi^{\overline{\mathbf{\Theta}}_{{\rm A}}}\left(\cdot\right)$ and $Q\left(\left.\cdot\right|\overline{\mathbf{\Theta}}_{{\rm C}}\right)$, which are clone from their online networks every a few steps. By feeding $\mathbf{o}\left(t+1\right)$ to the target networks, we can estimated the target Q value $Q\left(\left.t+1\right|\overline{\mathbf{\Theta}}_{{\rm C}}\right)$ which is a step forward for estimating the Q value and contributes to the critic loss $L\left(\mathbf{\Theta}_{{\rm C}}\right)$ together with $Q\left(\left.t\right|\mathbf{\Theta}_{{\rm C}}\right)$.

During the practical training, the mini-batch training combined with the experience replay is adopted to enhance the training stability. Specifically, the interaction data between the agent with the environment is stored in the replay buff $\varOmega=\left\{ \left\{ \mathbf{a}\left(t-1\right),\mathbf{o}\left(t\right),\mathbf{a}\left(t\right),r\left(t\right),\mathbf{o}\left(t+1\right)\right\} \right\}$. We randomly taken $M_{{\rm mini}}$ samples from $\varOmega$ as a mini-batch. Then, the critic loss in mean squared error (MSE) sense can be defined as
\begin{equation}\label{e20}
L\left(\mathbf{\Theta}_{{\rm C}}\right)=\frac{1}{M_{{\rm mini}}}\sum_{m=1}^{M_{{\rm mini}}}\left(y\left(t_{m}\right)-Q\left(\left.t_{m}\right|\mathbf{\Theta}_{{\rm C}}\right)\right)^{2},
\end{equation}
where 
\begin{equation}\label{e21}
y\left(t_{m}\right)=r\left(t\right)+\chi Q\left(\left.t_{m}+1\right|\overline{\mathbf{\Theta}}_{{\rm C}}\right).
\end{equation}
$t_{m}$ is the random sampling point in time. As such, $\mathbf{\Theta}_{{\rm C}}$ can be optimized by minimizing $L\left(\mathbf{\Theta}_{{\rm C}}\right)$ and the update gradient $\nabla_{\mathbf{\Theta}_{{\rm C}}}L\left(\mathbf{\Theta}_{{\rm C}}\right)$.

The actor network is supposed to output an optimal policy by maximizing the Q value.
\begin{algorithm}[!t]
\caption{RGRL Algorithm.}
\label{alg_PRGRL}
\begin{algorithmic}[1]
 \STATE  \textbf{Initialize:} Initialize $\mathbf{\Theta}_{{\rm A}}$, $\mathbf{\Theta}_{{\rm C}}$ randomly. Clone $\overline{\Theta}^{\textrm{A}}$, $\overline{\Theta}^{\textrm{C}}$.
\STATE \textbf{For} episode $=1,2,\cdots,\varPsi$ \textbf{do:}
\STATE \quad Initialize user sets $\mathcal{U}^{n},\forall n\in\mathcal{N}$ and $\mathbf{a}\left(0\right)$ randomly.
\STATE \quad \textbf{For} $t=1,2,\cdots,T$ \textbf{do:}
\STATE \quad \quad Observe the state $\mathbf{o}\left(t\right)$, and the reward $r\left(t\right)$ by \eqref{e15} \\
\STATE \quad \quad Update $\mathbf{o}\left(t\right)$ to $\mathbf{o}\left(t+1\right)$ under action $\mathbf{a}\left(t\right)$ by \eqref{e18}.
\STATE \quad \quad Store $\left\{ \mathbf{a}\left(t-1\right),\mathbf{o}\left(t\right),\mathbf{a}\left(t\right),r\left(t\right),\mathbf{o}\left(t+1\right)\right\}$ in $\Omega$.
\STATE \quad \quad Randomly sample a mini-batch data from $\Omega$.
\STATE \quad \quad Calculate $y\left(t_{m}\right)$ by \eqref{e21}. Then update $\mathbf{\Theta}_{{\rm C}}$ by \eqref{e20}\\
       \quad \quad and $\mathbf{\Theta}_{{\rm A}}$ by \eqref{e25}.
\STATE \quad \quad Soft-update the target actor/critic every $\phi$ steps:
\[\left\{ \begin{array}{c}
\overline{\mathbf{\Theta}}_{{\rm A}}\leftarrow\nu\mathbf{\Theta}_{{\rm A}}+\left(1-\nu\right)\overline{\mathbf{\Theta}}_{{\rm A}}\\
\overline{\mathbf{\Theta}}_{{\rm C}}\leftarrow\nu\mathbf{\Theta}_{{\rm C}}+\left(1-\nu\right)\overline{\mathbf{\Theta}}_{{\rm C}}
\end{array}\right.\]
\STATE \quad \textbf{End For}
\STATE \textbf{End For}
\end{algorithmic}
\end{algorithm}
Therefore, we can evaluate the current policy by defining a performance objective function
\begin{equation}
J_{\beta}\left(\pi\right)=\mathbb{E}_{s\sim\rho^{\beta}}\left[Q\left(\left.t\right|\mathbf{\Theta}_{{\rm C}}\right)\right],
\end{equation}
which estimates the expectation of $Q\left(\left.t\right|\mathbf{\Theta}_{{\rm C}}\right)$ under the state distribution $s\sim\rho^{\beta}$. Then, the policy gradient with respect to $\mathbf{\Theta}_{{\rm A}}$ can be given by
\begin{equation}
\begin{array}{l}
\nabla_{\mathbf{\Theta}_{{\rm A}}}J_{\beta}\left(\pi\right)=\\ \mathbb{E}_{s\sim\rho^{\beta}}\left[\nabla_{\mathbf{a}}Q\left(\left.t\right|\mathbf{\Theta}_{{\rm C}}\right)\cdot\nabla_{\mathbf{\Theta}_{{\rm A}}}\pi^{\mathbf{\Theta}_{{\rm A}}}\left(\mathbf{o}\left(t\right),\mathbf{a}\left(t-1\right)\right)\right].
\end{array}
\end{equation}
In practical, the mini-batch Monte Carlo sampling with a size of $I$ is adopted to yield a unbiased estimation of $\nabla_{\mathbf{\Theta}_{{\rm A}}}J_{\beta}\left(\pi\right)$.
\begin{equation}\label{e25}
\begin{array}{l}
\nabla_{\mathbf{\Theta}_{{\rm A}}}J_{\beta}\left(\pi\right)\approx\\ \frac{1}{I}\sum_{i=1}^{I}\left(\nabla_{\mathbf{a}}Q\left(\left.t_{i}\right|\mathbf{\Theta}_{{\rm C}}\right)\cdot\nabla_{\mathbf{\Theta}_{{\rm A}}}\pi^{\mathbf{\Theta}_{{\rm A}}}\left(\mathbf{o}\left(t_{i}\right),\mathbf{a}\left(t_{i}-1\right)\right)\right).
\end{array}
\end{equation}
The overall process of the proposed algorithm is summarized in {\bf{Algorithm \ref{alg_PRGRL}}}, where $\nu$ is coefficient of the soft-update.

\section{USE CASE DISCUSSION AND ANALYSIS}\label{sec_4}
In order to further verify the universal superiority of the proposed RGRL algorithm in solving the problem of hybrid resources allocation in MEC-assisted RAN slicing system oriented to heterogeneous service scenarios, this section discuss applications of the proposed RGRL algorithm in two classic use cases, i.e., single-node MEC-assisted RAN slicing scenario and non-cooperative multi-node MEC-assisted RAN slicing scenario.

\subsection{Use Case 1: Single-node MEC-Assisted RAN Slicing Scenario}

\begin{figure}[!t]
\centering
\includegraphics[width=0.25\textwidth, angle=90]{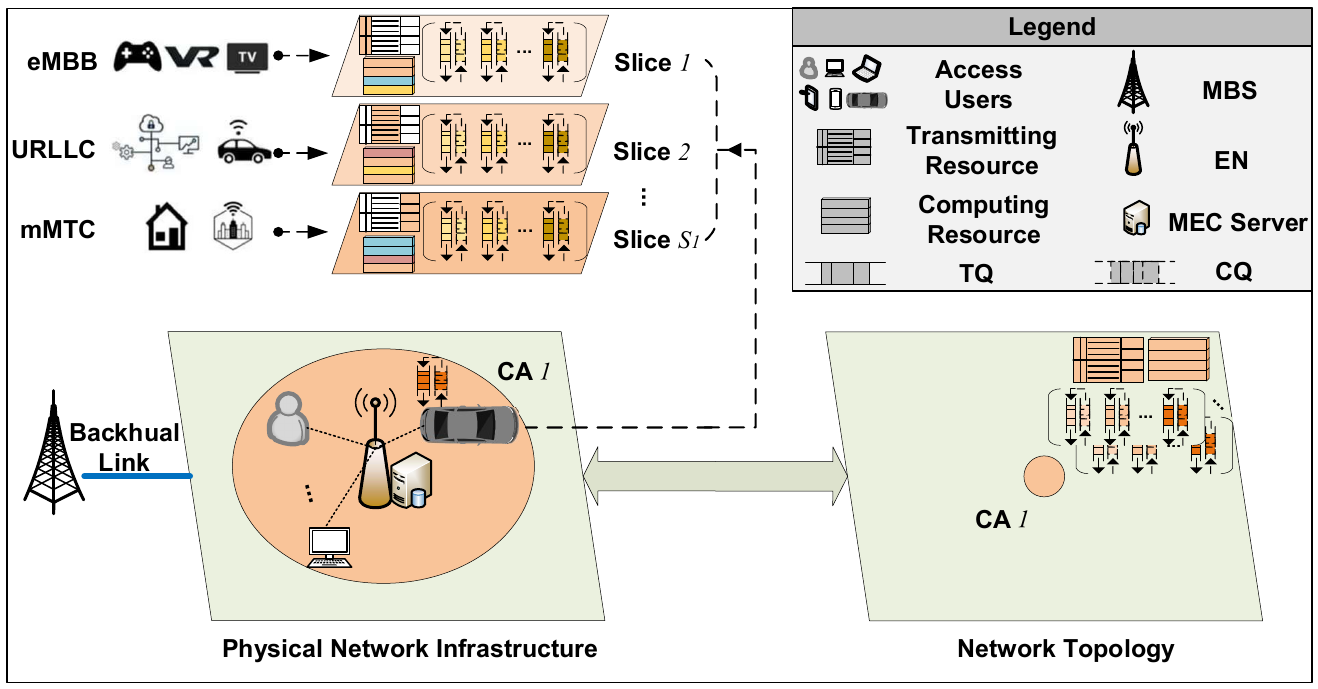}
\caption{Use case 1: Single-node MEC-assisted RAN slicing system model.}
\label{fig_6.3}
\end{figure}

As a simplest system, the single-node MEC-assisted RAN slicing system for the heterogeneous service scenario can quickly and effectively evaluate the universality and superiority of the proposed RGRL algorithm. Specifically, in this use case, the system includes a single MBS, a single EN, and $U$ AUs. Similarly, an MEC server with computing frequency of $C _ {\rm B}$ is deployed at the EN. The difference is that there is only one EN in this use case, and thus none of other ENs can be coordinated for cooperative computing. It is also assumed that the EN divides $S_1$ slices and has $Z_{\rm B}$ transmission RBs, which are used for the three heterogeneous services, i.e., eMBB, uRLLC and mMTC. It is not difficult to observer that the topological graph abstracted form this system framework of the use case 1 degenerates into a single-node trivial graph, as shown in Fig. \ref{fig_6.3}.

In this use case, the detial system models, such as task generation model, hybrid resource allocation model, queue model, and performance metric model, are similar to those of the aforementioned cooperative multi-node MEC network slicing system, where only need to set $N=1$. Therefore, the long-term average SSR maximization problem can be formulated as
\begin{subequations}\label{eq_6.27}
    \begin{align}
	\quad
    \underset{\mathcal{A}}{{\rm max}} \quad
    &\underset{T\rightarrow\infty}{{\rm lim}}\sum_{\tau=0}^{T}\mathbb{E}\left[\chi^{\tau}U_{{\rm avg}}\left(t+\tau\right)\right],\label{eq_6.27a}\\
    {\rm{s.t.}}\quad \quad
    & \sum_{j=1}^{N}c_{j}^{n}\left(t\right)=1,\forall n\in\mathcal{N}=\left\{ 1\right\} , \label{eq_6.27b}\\
    &\sum_{s=1}^{S_{n}}c^{n,s}\left(t\right)=1,\sum_{s=1}^{S_{n}}z^{n,s}\left(t\right)=1,\forall n\in\mathcal{N}=\left\{ 1\right\} ,\label{eq_6.27c}\\
    &\sum_{u\in\mathcal{U}_{s}^{n}}c^{n,s,u}\left(t\right)=1,\sum_{u\in\mathcal{U}_{s}^{n}}z^{n,s,u}\left(t\right)=1,\nonumber \\
    &\forall s\in\mathcal{S}_{n},\forall n\in\mathcal{N}=\left\{ 1\right\} ,\label{eq_6.27d}
    \end{align}
\end{subequations}
where $N=1,\mathcal{N}=\left\{1\right\}$. Similar to the problem \eqref{Pro_form}, we also convert the problem \eqref{eq_6.27} into a MDP problem, expressed as:

\subsubsection{State}
The state of the EN at time $t$ can be denoted as $\mathbf{o}^{1}\left(t\right)=\left[{\rm vec}\left\{ \mathcal{Q}_{u}^{1,s}\left(t\right)\right\} ,{\rm vec}\left\{ \mathcal{V}_{u}^{1,s}\left(t\right)\right\} ,h_{u}^{1,s}\left(t\right)\right]_{u\in\mathcal{U}_{s}^{1},s\in\mathcal{S}_{1}}$. The entire system state can be defined as $\ensuremath{\mathbf{o}\left(t\right)=\left[\mathbf{o}^{1}\left(t\right),\tilde{\mathbf{A}}\right]}$, where $\tilde{\mathbf{A}}=\left[1\right]$.

\subsubsection{Action}
The action of the single-node system can be denoted as $\ensuremath{\mathbf{a}\left(t\right)=\left[\mathbf{c}\left(t\right),\mathbf{z}\left(t\right)\right]}$, where $N=1$.

\subsubsection{State Transition}
Due to the existence of channel coefficients in the system state, the transition probability of the system state is still unavailable.

\subsubsection{Reward}
The instantaneous reward of a single-node system at time $t$ and the cumulative reward starting from time $t$ are defined as $r\left(t\right)=U_{{\rm avg}}\left(t\right)$ and $R\left(t\right)=\sum_{\tau=0}^{T}\chi^{\tau}r\left(t+\tau\right)$, respectively. It can be observed that these expressions are exactly the same as those of the cooperative multi-node system. Nevertheless, it should be noted that let $n=1$ in specific calculations.

Finally, let $N=1$ and $n\mathcal{\in N}=\left\{1\right\} $, then the proposed RGRL algorithm can be applied to solve the hybrid resource allocation problem \eqref{eq_6.27} formulated in this use case scenario.

\subsection{Use Case 2: Non-cooperative Multi-node MEC-Assisted RAN Slicing Scenario}
\begin{figure}[!t]
\centering
\includegraphics[width=0.28\textwidth, angle=90]{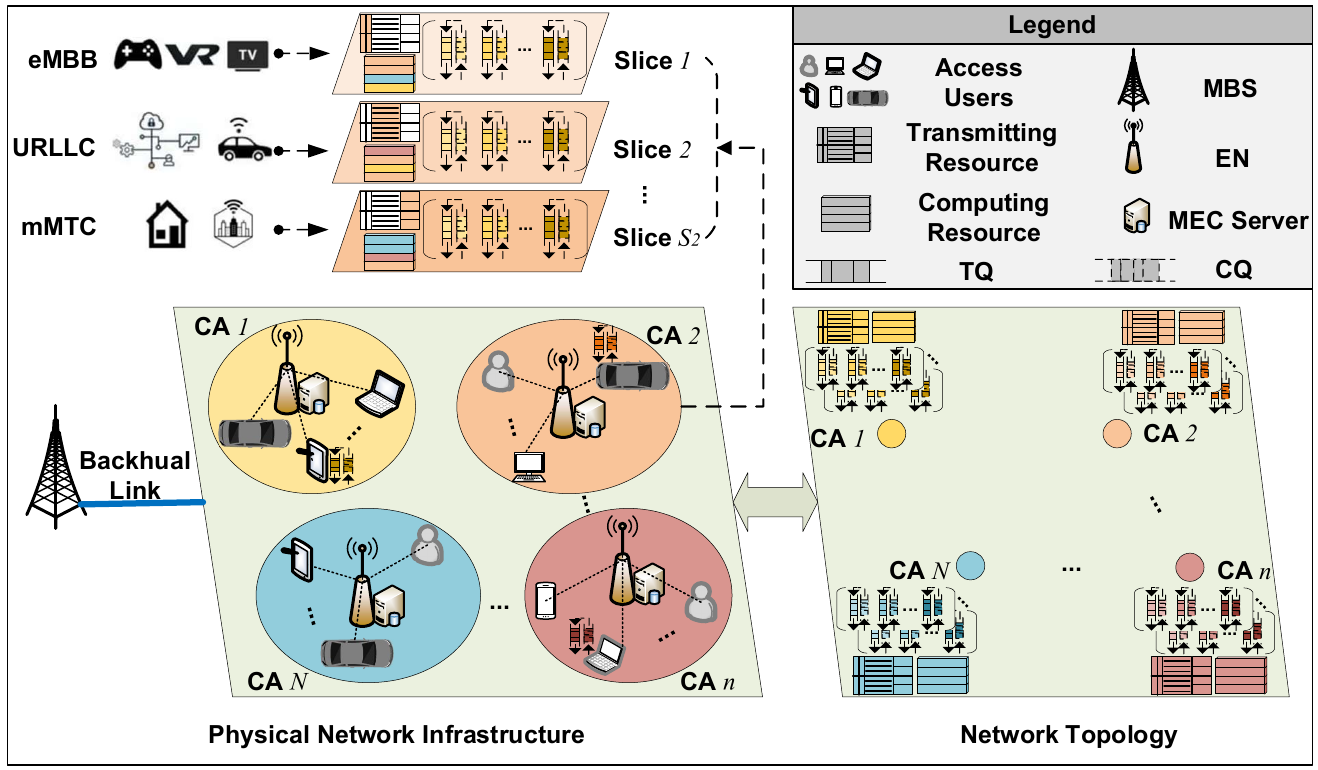}
\caption{Use case 2: Non-cooperative multi-node MEC-assisted RAN slicing system model.}
\label{fig_6.4}
\end{figure}

In this subsection, we discuss a non-cooperative multi-node MEC-assisted RAN slicing system for the heterogeneous service scenario, as shown in Fig. \ref{fig_6.4}. The configuration parameters are basically the same as those of the multi-node MEC network slicing system. Specifically, the system consists of one single MBS, $N$ ENs and $U$ AUs. Each EN is assumed to have $Z_{\rm B}$ transmission RBs and is equipped with a MEC server with computing frequency of $C_{\rm B}$. In addition, each EN can be divided into $S_n$ slices to provide users with three heterogeneous services, i.e., eMBB, uRLLC, and mMTC. The difference from the cooperative multi-node MEC network slicing system is that computing cooperations among ENs is no longer allowed in this use case 2. It is not difficult to find that the use case 2 is equivalent to the special case where $ A_{\rm max}=0$ in the cooperative multi-node MEC network slicing system. Correspondingly, the topology graph of the system framework in this use case 2 can be abstracted as a null graph, as shown in Fig. \ref{fig_6.4}. As such, the self-connection weighted adjacency matrix of the topology degenerates into an identity matrix of order $N$, expressed as $\tilde{\mathbf{A}}=\mathbf{I}_{N}$.

The construction of the specific task generation model, queue model and other system models in this use case 2 is still the same as that of the multi-node MEC network slicing system. Therefore, for this non-cooperative multi-node MEC network slicing system, the long-term cumulative average SSR maximization problem under the hybrid resource allocation policy can be defined as
\begin{subequations}\label{eq_6.28}
    \begin{align}
	\quad
    \underset{\mathcal{A}}{{\rm max}} \quad
    &\underset{T\rightarrow\infty}{{\rm lim}}\sum_{\tau=0}^{T}\mathbb{E}\left[\chi^{\tau}U_{{\rm avg}}\left(t+\tau\right)\right],\label{eq_6.28a}\\
    {\rm{s.t.}}\quad \quad
    & c_{j}^{n}\left(t\right)=\begin{cases}
1, & j=n\\
0, & j\neq n
\end{cases},\forall n,j\in\mathcal{N}, \label{eq_6.28b}\\
    &\sum_{s=1}^{S_{n}}c^{n,s}\left(t\right)=1,\sum_{s=1}^{S_{n}}z^{n,s}\left(t\right)=1,\forall n\in\mathcal{N},\label{eq_6.28c}\\
    &\sum_{u\in\mathcal{U}_{s}^{n}}c^{n,s,u}\left(t\right)=1,\sum_{u\in\mathcal{U}_{s}^{n}}z^{n,s,u}\left(t\right)=1,\nonumber \\
    &\forall s\in\mathcal{S}_{n},\forall n\in\mathcal{N},\label{eq_6.28d}
    \end{align}
\end{subequations}

Likewise, in order to apply the proposed RGRL algorithm to solve the hybrid policy optimization problem \eqref{eq_6.28}, it is necessary to transform the problem \eqref{eq_6.28} into a MDP problem, formulated as follows.

\subsubsection{State}
At time $t$, the state of EN $n$ can be denoted as $\mathbf{o}^{n}\left(t\right)=\left[{\rm vec}\left\{ \mathcal{Q}_{u}^{n,s}\left(t\right)\right\} ,{\rm vec}\left\{ \mathcal{V}_{u}^{n,s}\left(t\right)\right\} ,h_{u}^{n,s}\left(t\right)\right]_{u\in\mathcal{U}_{s}^{n},s\in\mathcal{S}_{n}}$. Then, the entire system state can be defined as $\mathbf{o}\left(t\right)=\left[\mathbf{o}^{1}\left(t\right),\cdots,\mathbf{o}^{N}\left(t\right),\tilde{\mathbf{A}}\right]$, where $\tilde{\mathbf{A}}=\mathbf{I}_{N}$.

\subsubsection{Action}
The resource allocation action of the system at time $t$ can be expressed as $\mathbf{a}\left(t\right)=\left[\mathbf{c}\left(t\right),\mathbf{z}\left(t\right)\right]$. In particular, when calculating the action of computing resource allocation, we have 
\begin{equation}\label{eq_6.29}
c_{j}^{n}\left(t\right)=\begin{cases}
1, & j=n\\
0, & j\neq n
\end{cases},\forall n,j\in\mathcal{N}
\end{equation}

\subsubsection{State Transition}
The state transition probability is still unavailable due to the existence of channel coefficients in the system state.

\subsubsection{Reward}
For problem \eqref{eq_6.28}, the instantaneous reward at time $t$ and the cumulative reward starting from time $t$ can be respectively defined as $r\left(t\right)=U_{{\rm avg}}\left(t\right)$ and $R\left(t\right)=\sum_{\tau=0}^{T}\chi^{\tau}r\left(t+\tau\right)$.

Finally, the proposed RGRL algorithm can be directly adapted to this use case scenario to learn the optimal hybrid policy.

\begin{table*}[!hbpt]
\centering
\caption{Simulation Parameters} \label{simu_setup}
\begin{tabular}{cc|cccccc}
\toprule[1pt]
\multicolumn{2}{c|}{Service}               & \multicolumn{2}{c}{eMBB}                                            & \multicolumn{2}{c}{mMTC}                                      & \multicolumn{2}{c}{uRLLC}                  \\ \hline
\multicolumn{2}{c|}{$U_{\rm mini}$}        & \multicolumn{2}{c}{3}                                               & \multicolumn{2}{c}{50}                                        & \multicolumn{2}{c}{10}                     \\ \hline
\multicolumn{2}{c|}{$l_{{\rm req}}^{n,s}$} & \multicolumn{2}{c}{0.05 s}                                           & \multicolumn{2}{c}{0.02 s}                                     & \multicolumn{2}{c}{0.001 s}                 \\ \hline
\multicolumn{2}{c|}{$\kappa_{u}^{n,s}$}   & \multicolumn{2}{c}{0.6$\sim$0.8}                                   & \multicolumn{2}{c}{0.4$\sim$0.6}                             & \multicolumn{2}{c}{0.8$\sim$1.0}          \\ \hline
\multicolumn{2}{c|}{$\zeta_{u}^{n,s}$}     & \multicolumn{2}{c}{5$\sim$10}                                      & \multicolumn{2}{c}{5$\sim$10}                                & \multicolumn{2}{c}{5$\sim$10}             \\ \hline
\multicolumn{2}{c|}{$\delta_{u}^{n,s}$}    & \multicolumn{2}{c}{[0.1,\;0.3] MB}                                & \multicolumn{2}{c}{125 B}                                      & \multicolumn{2}{c}{[10,\;300] B}          \\ \hline\hline
\multicolumn{1}{c|}{$C_{\rm B}$} & 10{\em GHz}   & \multicolumn{1}{c|}{$Z_{\rm B}$} & \multicolumn{1}{c|}{10}          & \multicolumn{1}{c|}{$P_{\rm B}$} & \multicolumn{1}{c|}{11{\em dBm}} & \multicolumn{1}{c|}{$W_{\rm B}$} & 0.18{\em MHz} \\ \hline
\multicolumn{1}{c|}{$N_0$}       & -204dBm & \multicolumn{1}{c|}{$\varpi$}    & \multicolumn{1}{c|}{15} & \multicolumn{1}{c|}{$\nabla t$}  & \multicolumn{1}{c|}{1{\em s}}    & \multicolumn{1}{c|}{$T$}         & 100     \\ 
\bottomrule[1pt]
\end{tabular}
\end{table*}
\section{NUMERICAL SIMULATIONS}\label{sec_5}

In this section, we discuss and evaluate the advantages and universality of the proposed RGRL algorithm from aforementioned three application scenarios. In simulations, we set $d_{0}=200\;\textrm{m}$, $d_{r}=100\;\textrm{m}$, $d_{min}=10\;\textrm{m}$, $d_{max}=100\;\textrm{m}$, and $\beta=2$. Moreover, $S_n$ is randomly integer sampled in the interval 3 to 6 while ensuring the number of slices for each type service is not less than 1. According to $\mathcal{U}_{s}^{n}\notin\emptyset,\forall n\in\mathcal{N},\forall s\in\mathcal{S}_{n}$, users are randomly assigned to each slice of ENs according $\mathcal{U}_{s}^{n}\notin\emptyset,\forall n,s$. Let $U_{\rm mini}$ be the minimum number of NS users. Other parameters are detailed in Table \ref{simu_setup} \cite{MeiJ21,Navarro20,GSA2017}. Note that, in following simulations, all the ENs will be randomly redistributed in the region when the value of $N$ or $U$, or the simulation scenario changes.

We compare the proposed algorithm with baseline algorithms including three commonly used RL algorithm for the RA in RAN slicing, i.e., dense layer based RL (Dense-RL) \cite{SinhaS20} which is the most common RL algorithm, GCN layer based RL (GCN-RL) \cite{JiangJ20} which can capture system space characteristics, and long short-term memory (LSTM) layer based RL (LSTM-RL) \cite{LiX15} which includes the tradition time recurrent structure to capture system time-dimensional characteristics, as well as a random algorithm in which a random action is executed regardless of the current state. Moreover, we also set a dense-based recurrent RL (Dense-RRL) baseline algrothm for further comparisons by embedding the proposed time recurrent architecture into the Dense-RL algorithm. For fairness, the DDPG framework is adopted as the basic framework of all those RL algorithms. Adam optimizer \cite{Kingma2014Adam} is used to optimize all the neural network parameters with a start learning rate of ${10^{ - 4}}$. In addition, for each scenario, we first train all involved algorithms and then performed the performance evaluations through 500 consecutive testing episodes. 

\begin{figure}[!t]
\centering
\includegraphics[width=0.41\textwidth]{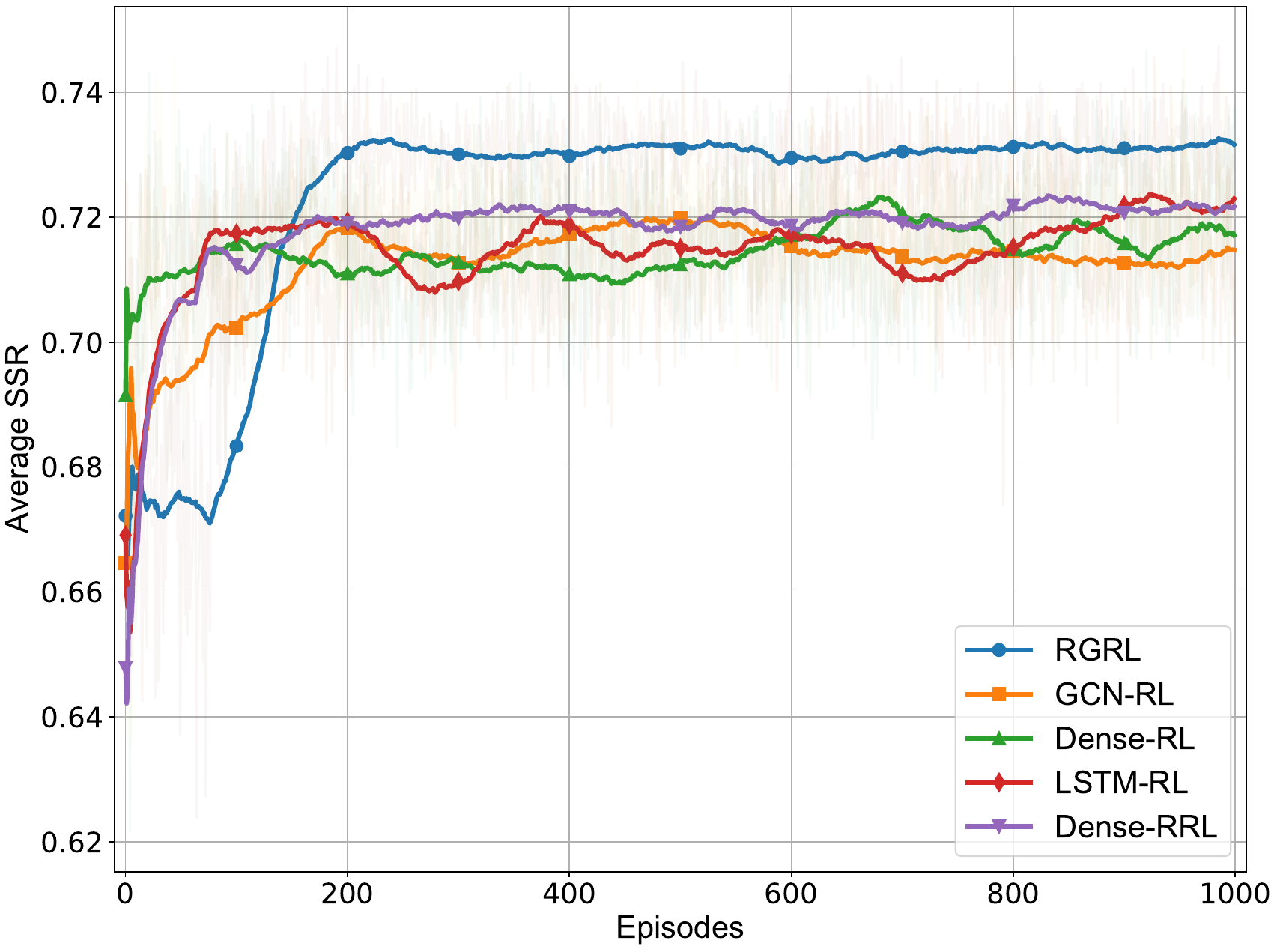}
\caption{Convergence behaviors of the  proposed RGRL algorithm. ($N=8$, $U=650$, $C_{\rm B}=10 \; {\rm GHz}$, $Z_{\rm B}=10$)}
\label{fig_6.4_5}
\end{figure}

\subsection{Simulation Results Under The Major Investigated Scenario}

\begin{figure}[!tbp]
\centering
\includegraphics[width=0.4\textwidth]{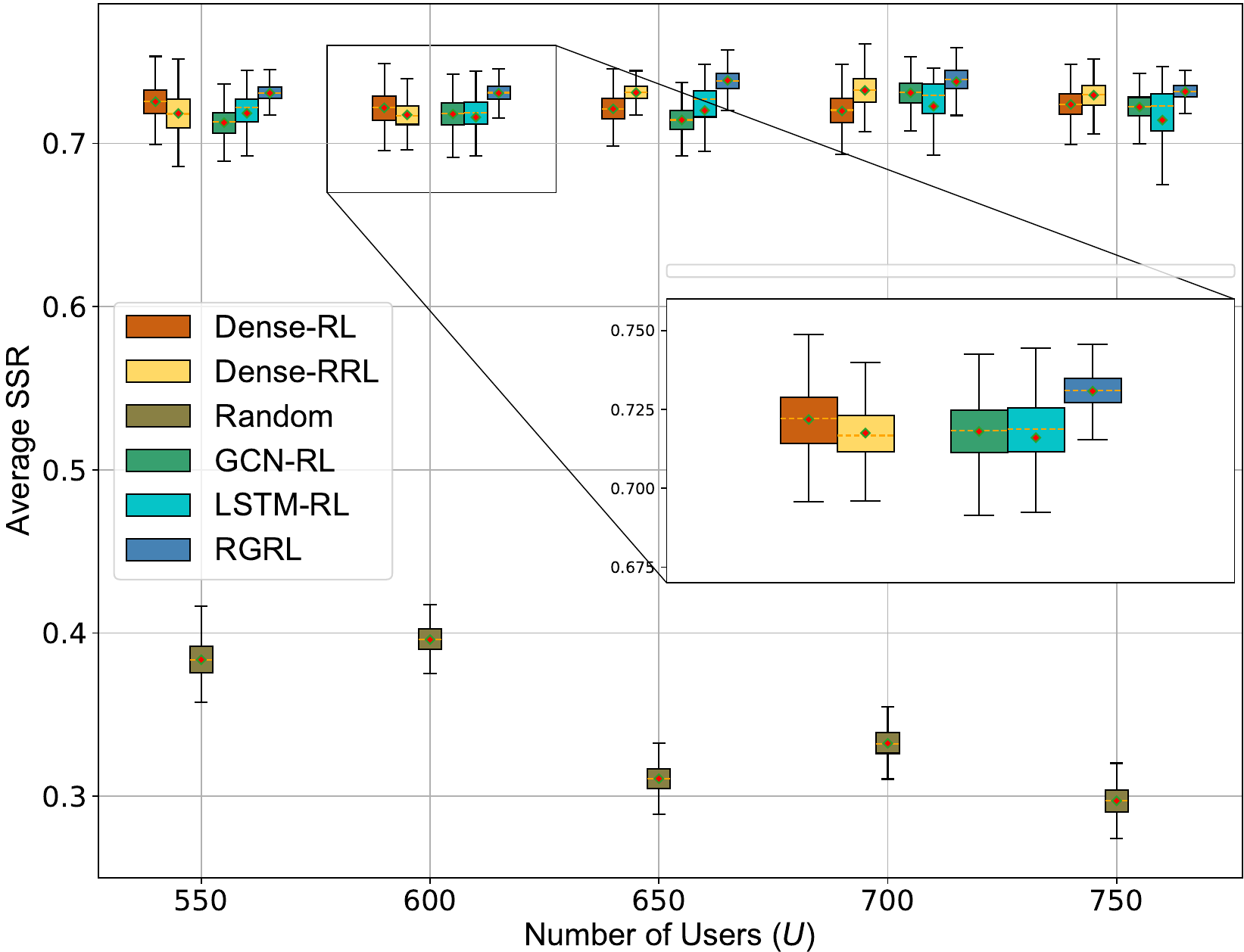}
\caption{Performance comparisons of the algorithms on different number of users under the major investigated scenario. ($N=8$, $C_{\rm B}=10 \; {\rm GHz}$, $Z_{\rm B}=10$)}
\label{numusers_compa}
\end{figure}

In this subsection, we evaluate the proposed RGRL algorithm under the cooperative multi-node MEC-assisted RAN slicing scenario, where $A_{\rm max}=3$. From the perspective of convergent behavior of the proposed RGRL algorithm, the training processes are illustrated in Fig. \ref{fig_6.4_5}. From Fig. \ref{fig_6.4_5}, we can observe that the proposed RGRL algorithm can converge to a stable level around 400 training episodes. Moreover, Fig. \ref{fig_6.4_5} implies that the RGRL algorithm is significantly superior to other baseline algorithm in terms of the convergent average SSR performance. In addition, it can be found from the volatility of each convergence curve in Fig. \ref{fig_6.4_5} that the proposed RGRL algorithm maintains a more stable convergence performance than all the baseline algorithms. The performance comparisons among all the algorithms versus the number of users $U$ are presented in Fig. \ref{numusers_compa}. Note that, the algorithm stability can be measured by the variance of the performance data, which is reflected on the box length. 
\begin{figure}[!tbp]
\centering
\includegraphics[width=0.4\textwidth]{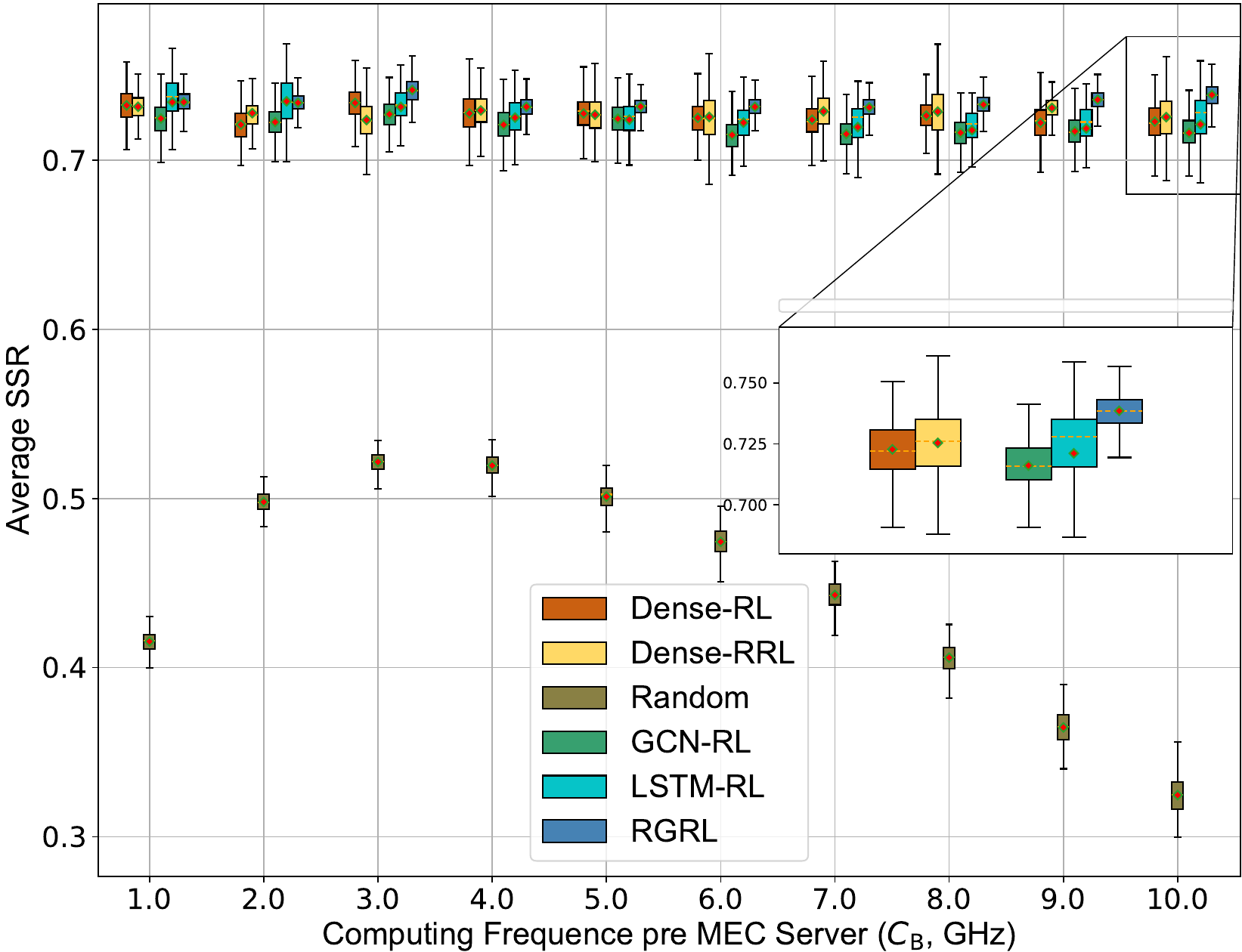}
\caption{Performance comparisons of the algorithms on different computing frequencies under the major investigated scenario. ($N=7$, $U=525$, $Z_{\rm B}=10$)}
\label{numcf_compa}
\end{figure}

It can be seen from Fig. \ref{numusers_compa} that the proposed RGRL algorithm outperforms all the baseline algorithms in terms of average SSR as well as stability regardless of the value of $U$. More specifically, we can observe from Fig. \ref{numusers_compa} that the performance of the Dense-RL, the GCN-RL and the LSTM-RL is comparable, which demonstrates that simply changing the type of neural network layer cannot improve the system performance effectively. Furthermore, the performance superiority of the proposed RGRL algorithm over the LSTM-RL algorithm indicates that the proposed time recurrent reinforcement learning framework is superior to the classical recurrent framework adopted in the recurrent neural network. In Fig. \ref{numcf_compa}, we provide the performance comparisons with respect to the computing capability $C_{\rm B}$ from 1 GHz to 10 GHz, while $N=7$ and $U=525$. It can be found that the average SSR and the stability of the proposed RGRL algorithm are both superior to other baseline algorithms under different $C_{\rm B}$. In particular, the average SSR advantage of the proposed RGRL becomes more significant as $C_{\rm B}$ increases. This implies proposed RGRL algorithm is more competitive with regard to the average SSR especially when the computing resource of the MEC server is more abundant.

\begin{figure}[!t]
\centering
\includegraphics[width=0.4\textwidth]{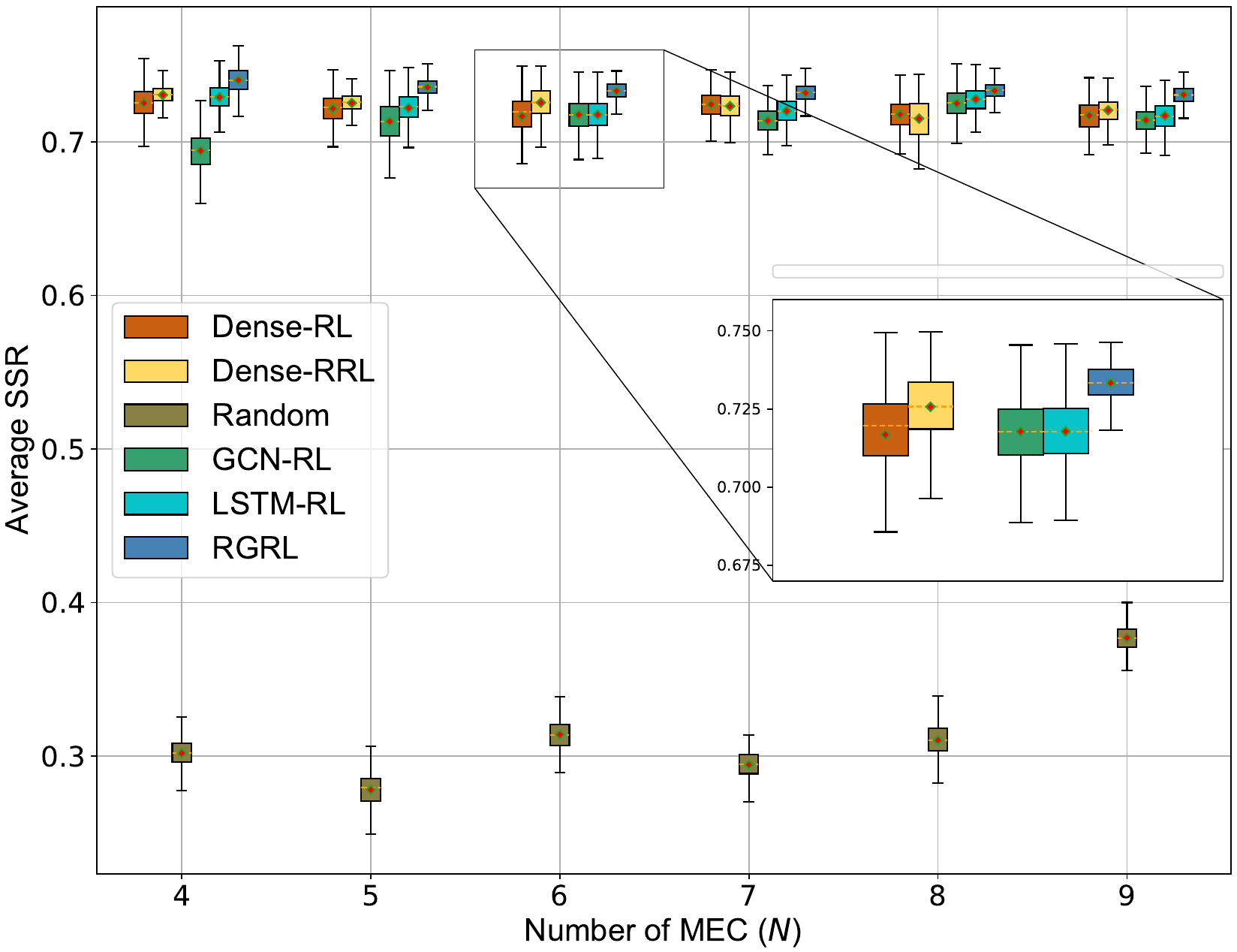}
\caption{Performance comparisons of the algorithms on different number of ENs under the major investigated scenario. ($U=650$, $C_{\rm B}=10 \; {\rm GHz}$, $Z_{\rm B}=10$)}
\label{fig_6.7}
\end{figure}
\begin{figure}[!t]
\centering
\includegraphics[width=0.4\textwidth]{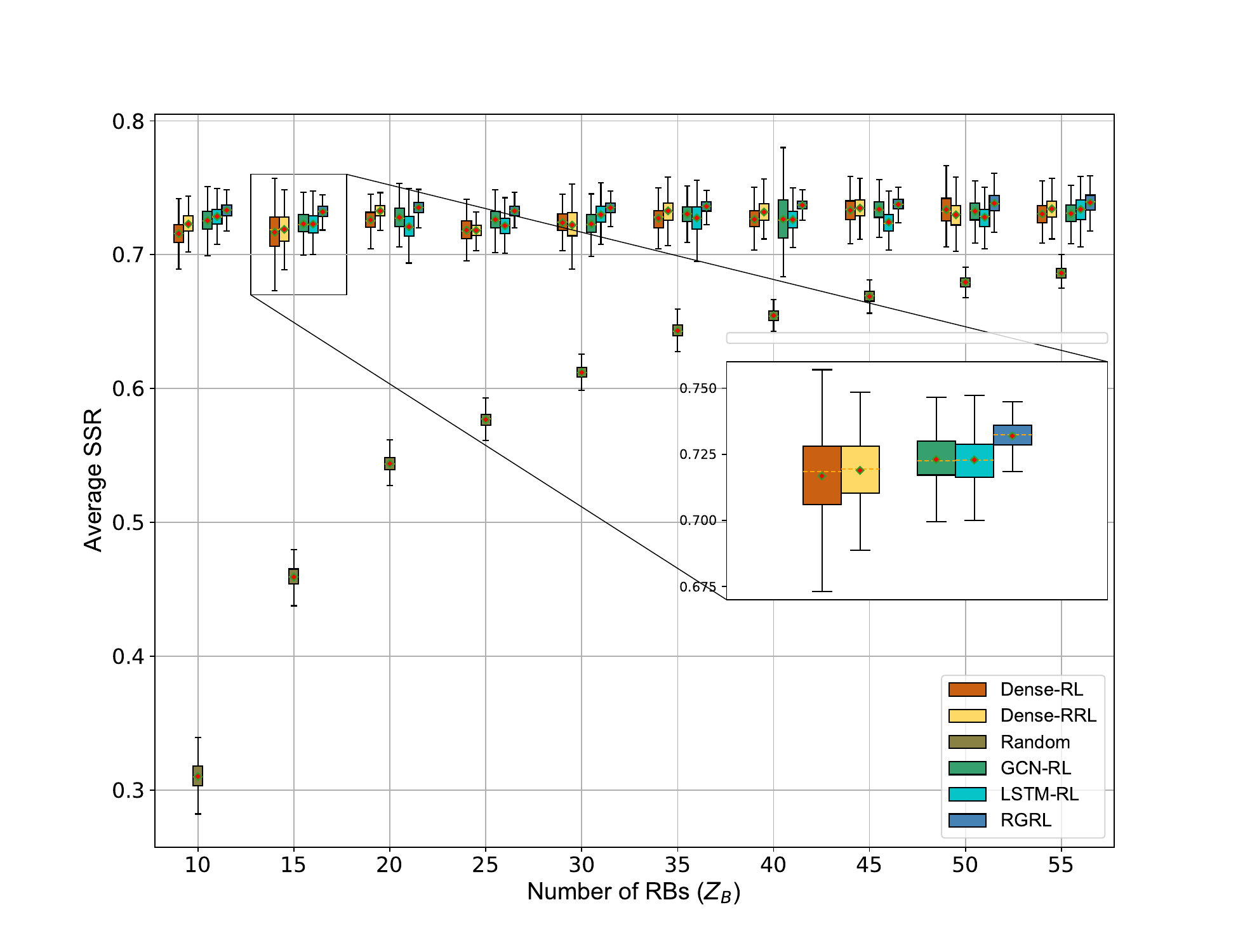}
\caption{Performance comparisons of the algorithms on different number of RBs under the major investigated scenario. ($N=8$, $U=650$, $C_{\rm B}=10 \; {\rm GHz}$)}
\label{fig_6.8}
\end{figure}

\begin{figure}[!t]
\centering
\subfigure[]{\includegraphics[width=0.39\textwidth]{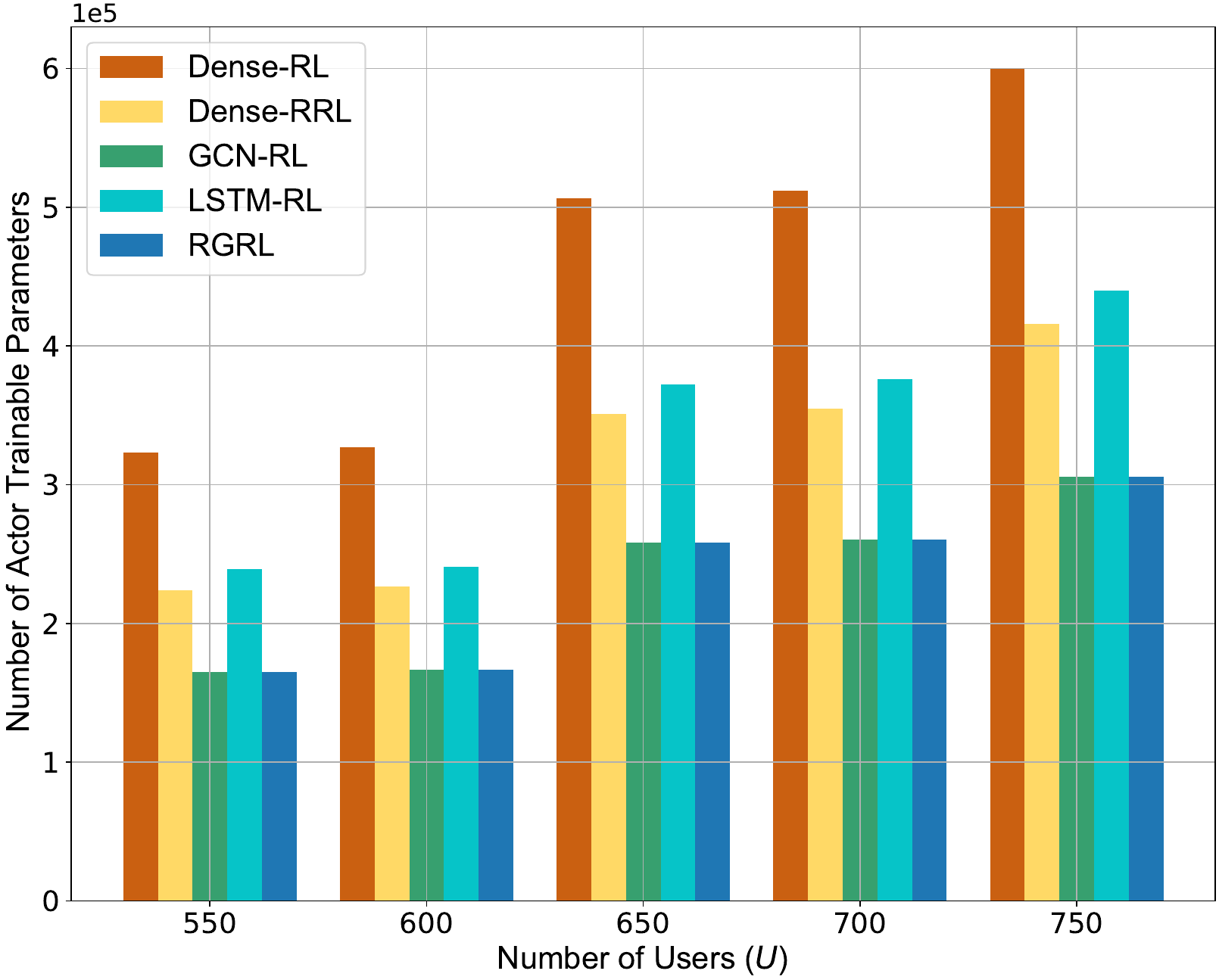}}
\subfigure[]{\includegraphics[width=0.39\textwidth]{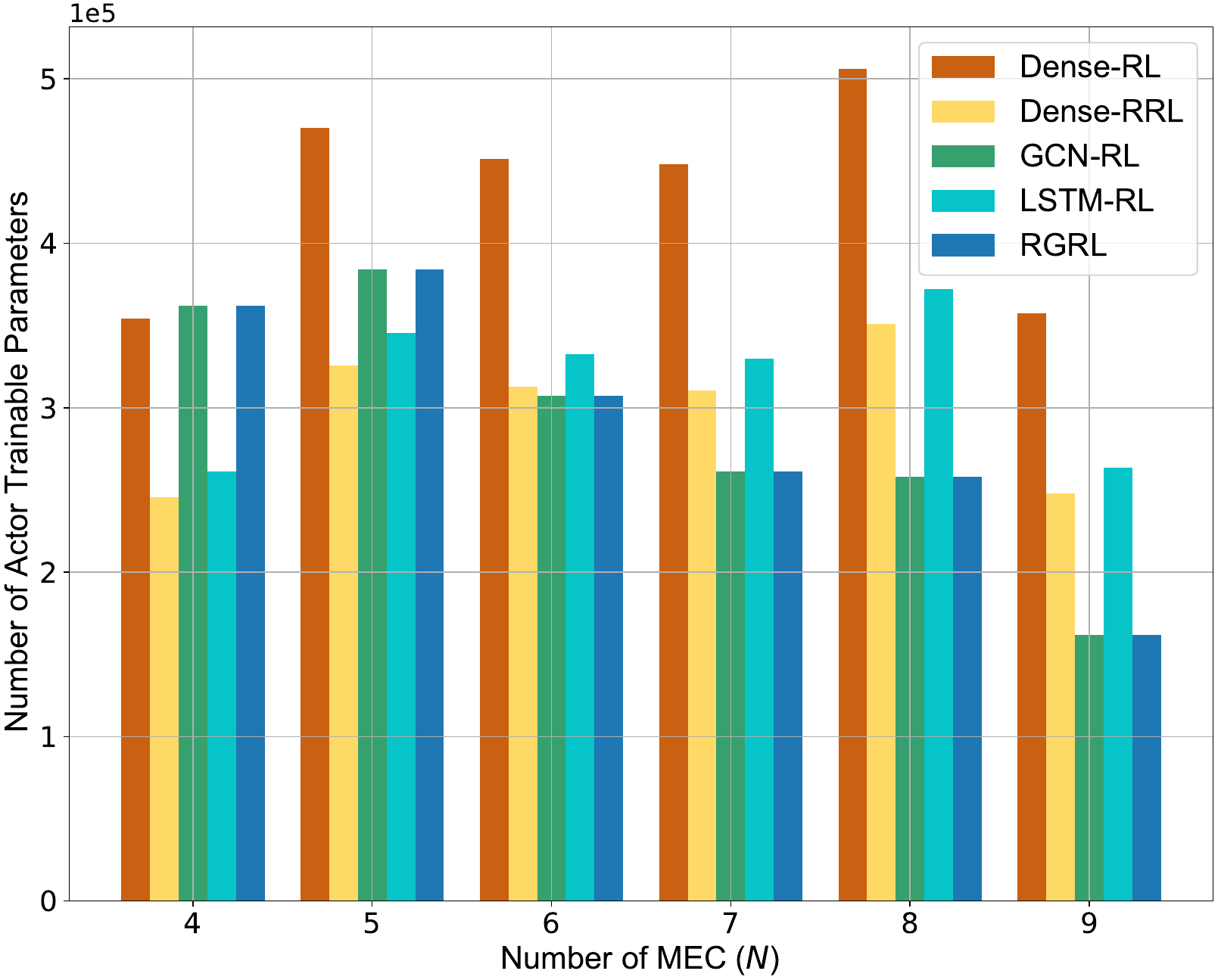}}
\subfigure[]{\includegraphics[width=0.39\textwidth]{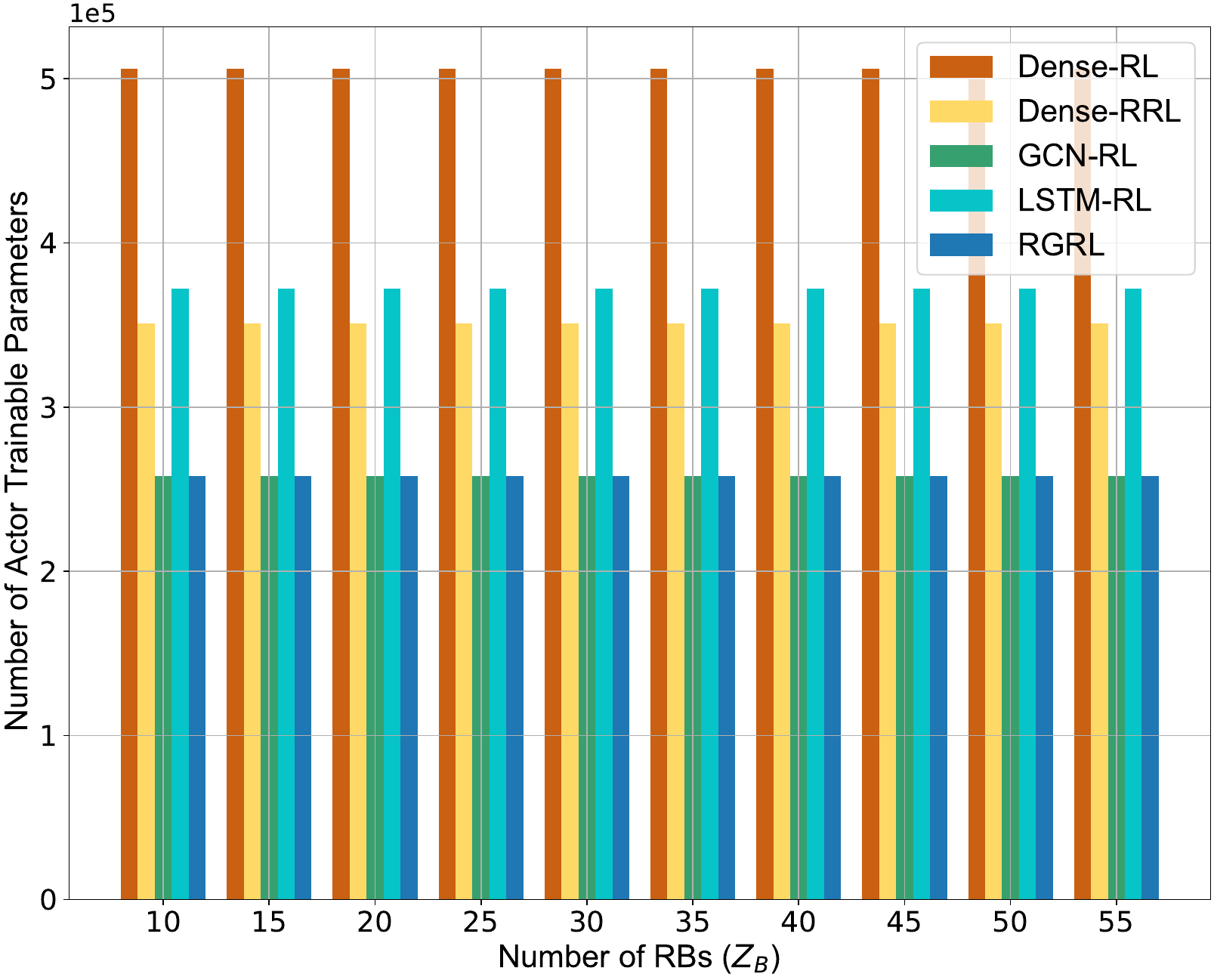}}
\caption{Network complexity evaluations of the algorithms under the major investigated scenario. (a) Evaluations on different number of users. ($N=8$, $Z_{\rm B}=10$) (b) Evaluations on different number of ENs. ($U=650$, $Z_{\rm B}=10$) (c) Evaluations on different number of RBs. ($N=8$,$U=650$)}
\label{complexity_compar}
\end{figure}
Fig. \ref{fig_6.7} provides the performance evaluation of the algorithms under different numbers of ENs. We can observe from Fig. \ref{fig_6.7} that the proposed RGRL algorithm outperforms all the baseline algorithms regardless of the value of $N$. Moreover, according to the length of the boxes in Fig. \ref{fig_6.7}, it can be seen that the the proposed RGRL algorithm is more stable than all the baseline algorithms. We also readily observe from Fig. \ref{fig_6.7} that both the GCN-RL baseline algorithm and the Dense-RRL baseline algorithm perform poorly. This is because the GCN-RL baseline algorithm only exploits GCN to capture spatial information from the cooperative multi-nodes system but ignores the contextual relevance of the system in time dimension. Although the Dense-RRL baseline algorithm deploys the proposed time recurrent reinforcement learning architecture, it is difficult for the fully connected neural network to extract spatial information from the multi-node system. Finally, the performance comparisons between the proposed RGRL algorithm and all the baseline algorithms versus the number of transmission RBs $Z_{\rm B}$ are presented in Fig. \ref{fig_6.8}. It can be found from Fig. \ref{fig_6.8} that the proposed RGRL algorithm still outperform all the baseline algorithms in terms of average SSR. In addition, we can observe from Fig. \ref{fig_6.8} that all the algorithms can never approach 100\% average SSR with the increase of $Z_{\rm B}$, which is due to the constraint of the calculation frequency $C_{ \rm B}$.

We take a step forward and evaluate the network complexity of all algorithms in terms of a commonly used metric, i.e., number of trainable parameters. Note that all the algorithms are trained offline and executed online, where online execution determines the service performance of the communication system and only actors participate in the online execution of each algorithm as the outputter of the hybrid RA policy. Therefore, we focus on the network complexity of the actor network for each algorithm. Specifically, network complexity evaluations of all the algorithms under the major investigated scenario are illustrated in Fig. \ref{complexity_compar}, where the numbers of actor trainable parameters are given by the actual statistics of the {\em Python} library. We can observe from Fig. \ref{complexity_compar} (a) and (c) that the parameters amount of the proposed RGRL algorithm is less than all the other baseline algorithms regardless of the value of $U$ and $Z_{\rm B}$. From Fig. \ref{complexity_compar} (b), we can find that the network complexity of the proposed RGRL algorithm is still competitive with other baseline algorithms under different value of $N$, especially in the case of large $N$. Moreover, we can find from Fig. \ref{complexity_compar} that bars of the RGRL algorithm and the GCN-RL algorithm are of equal height, which indicates that the designed time recurrent structure will not increase the network complexity of RL algorithms. By comparing bars of the LSTM-RL algorithm and Dense-RRL algorithm in Fig. \ref{complexity_compar}, we can further find that the designed time recurrent structure is superiority to the traditional LSTM recurrent structure in terms of network complexity. Finally, through discussions of Fig. \ref{fig_6.4_5} to Fig. \ref{complexity_compar}, we can conclude that the proposed RGRL lgorithm can achieve higher and more stable average SSR performance while reducing network complexity.

\subsection{Simulation Results Under The Use Case 1}

\begin{figure}[!t]
\centering
\includegraphics[width=0.4\textwidth]{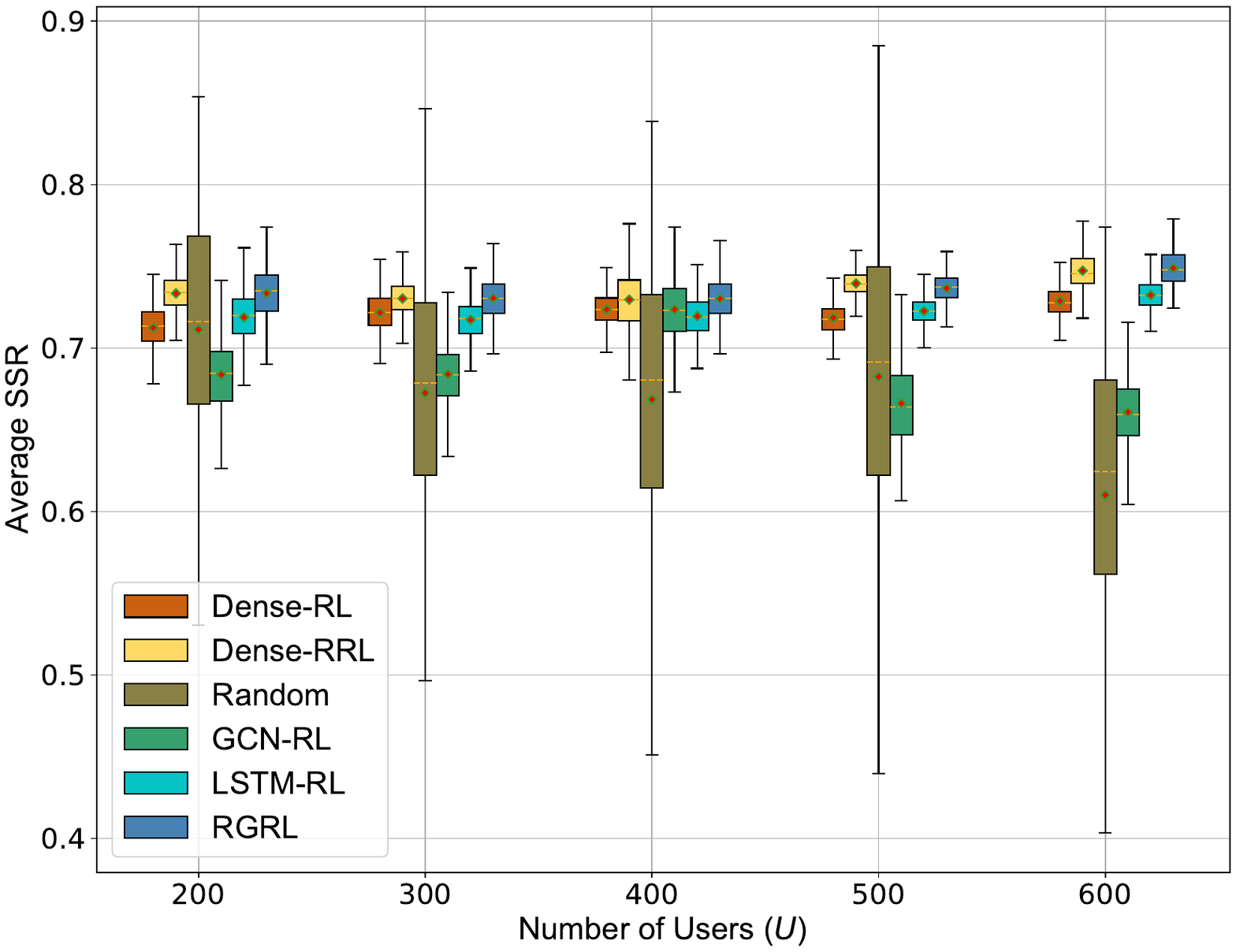}
\caption{Performance comparisons of the algorithms on different number of users under the use case 1. ($N=1$, $C_{\rm B}=10 \; {\rm GHz}$, $Z_{\rm B}=10$)}
\label{fig_6.9}
\end{figure}

\begin{figure}[!t]
\centering
\includegraphics[width=0.4\textwidth]{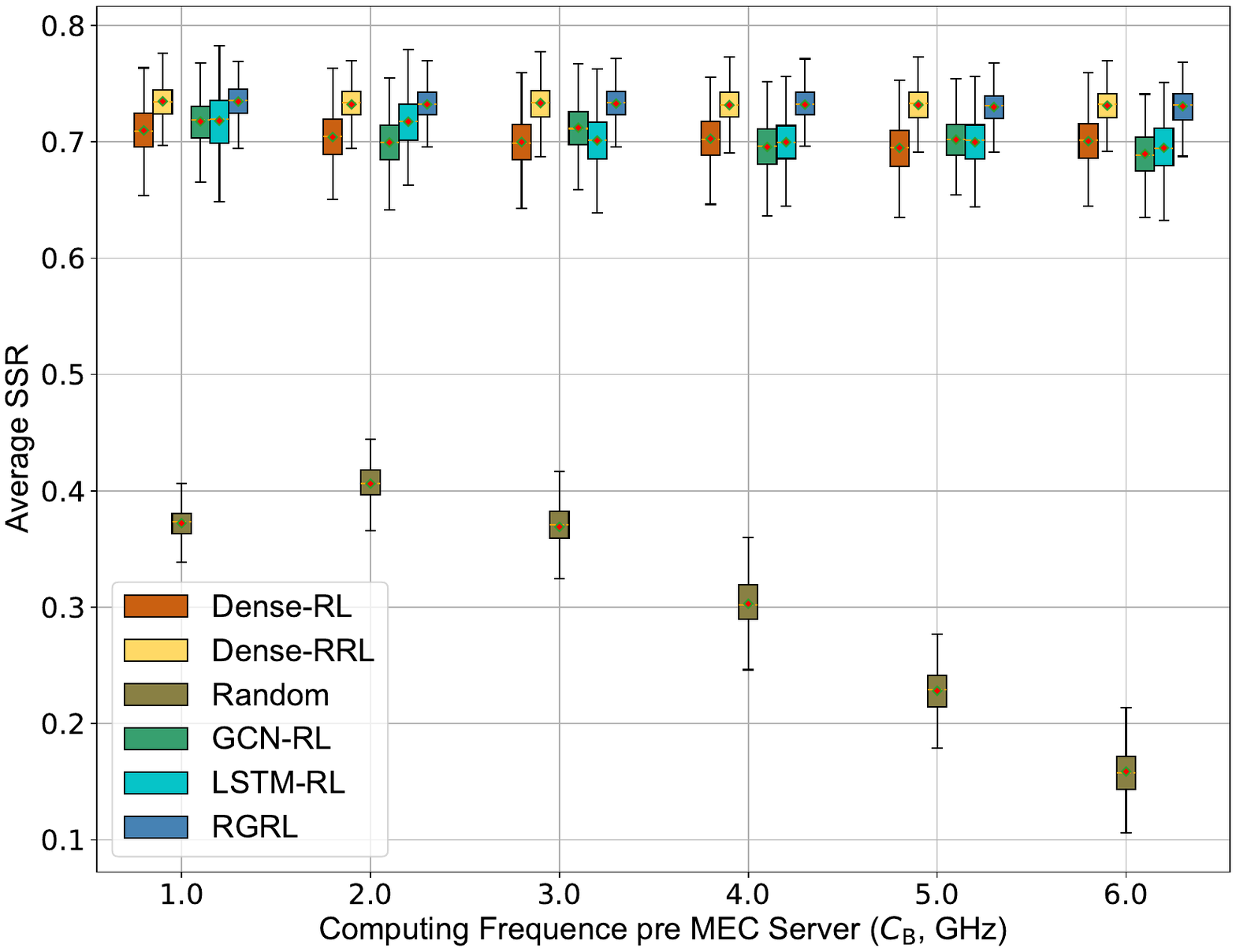}
\caption{Performance comparisons of the algorithms on different computing frequencies under the use case 1. ($N=1$, $U=100$, $Z_{\rm B}=10$)}
\label{fig_6.10}
\vspace{2.9mm}
\end{figure}

In this subsection, we evaluate the proposed RGRL algorithm under the single-node MEC-assisted RAN slicing scenario, where $N=1$. Fig. \ref{fig_6.9} provides the performance evaluation of the algorithm under different number of users $U$. Fig. \ref{fig_6.9} shows that proposed RGRL algorithm has comparable performance to the baseline Dense-RRL algorithm in terms of the average SSR. The reason is that this investigated use case scenario only consists of one single EN and thus contains no spatial information. As such, the fully connected neural network and the GCN have the same effect. Furthermore, we observe from Fig. \ref{fig_6.9} that the proposed RGRL algorithm and the Dense-RRL algorithm outperform all the other baseline algorithms regardless of the value of $U$, which demonstrates the superiority of the proposed time recurrent reinforcement learning framework. In addition, it can be seen that the baseline Random algorithm Fig. \ref{fig_6.9} achieves a decent mean performance on the average SSR while with an intolerable variance performance. The reason is that current computing and transmission resources are sufficient for the single-node system, thus the Random algorithm can obtain a decent mean performance under random resource allocation actions, while cannot maintain a stable average SSR due to the lack of accurate and adaptive resource management. The performance comparison among the proposed RGRL algorithm and all the baseline algorithms versus the computing frequencies $C_{\rm B}$ from 1 GHz to 6 GHz is presented in Fig. \ref{fig_6.10}. It can be seen from Fig. \ref{fig_6.10} that the proposed RGRL algorithm is still basically the same with the Dense-RRL algorithm in terms of average SSR under different $C_{\rm B}$, but outperforms all the other baseline algorithms. This further implies superiority of the proposed time recurrent reinforcement learning framework.

\begin{figure}[!t]
\centering
\includegraphics[width=0.4\textwidth]{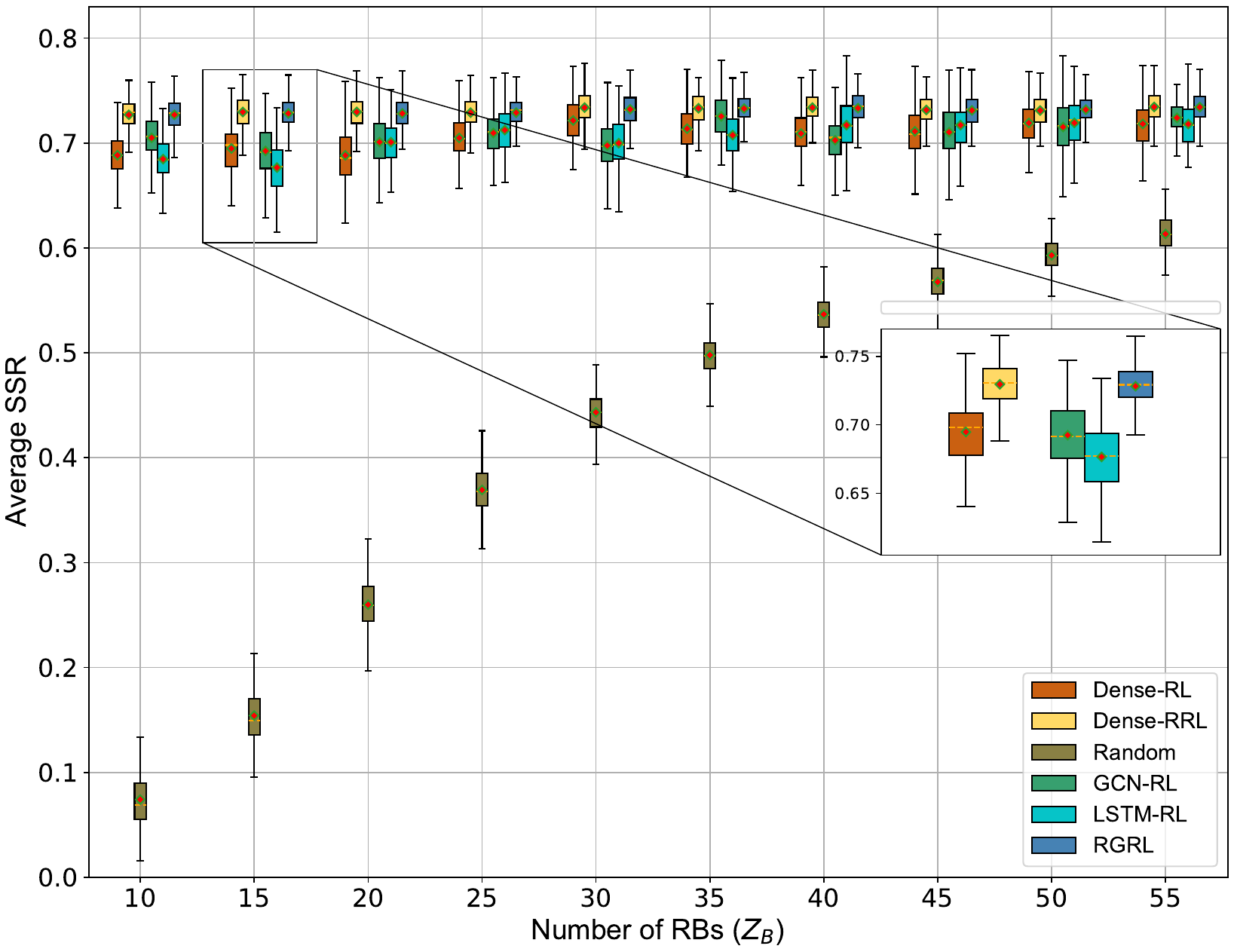}
\caption{Performance comparisons of the algorithms on different number of RBs under the use case 1. ($N=1$, $U=100$, $C_{\rm B}=10 \; {\rm GHz}$)}
\label{fig_6.11}
\end{figure}

Fig. \ref{fig_6.11} further provides the performance evaluation of the proposed RGRL algorithm on different number of transmission RBs $Z _ {\rm B}$. Similarly, Fig. \ref{fig_6.11} shows that the average SSR performance of the proposed RGRL algorithm and the baseline Dense-RRL algorithm still maintain basically consistent while consistently outperform other baseline algorithms. This is consistent with the aforementioned inferences about the lack of spatial features in the single-node system. In addition, by combining Fig. \ref{fig_6.8} and Fig. \ref{fig_6.11}, we can observe that unlimited increase of transmission RBs in the single-node system cannot promote the system to obtain 100\% average SSR, because the computing frequency is another constraint on the system performance.

\begin{figure}[!t]
\centering
\includegraphics[width=0.4\textwidth]{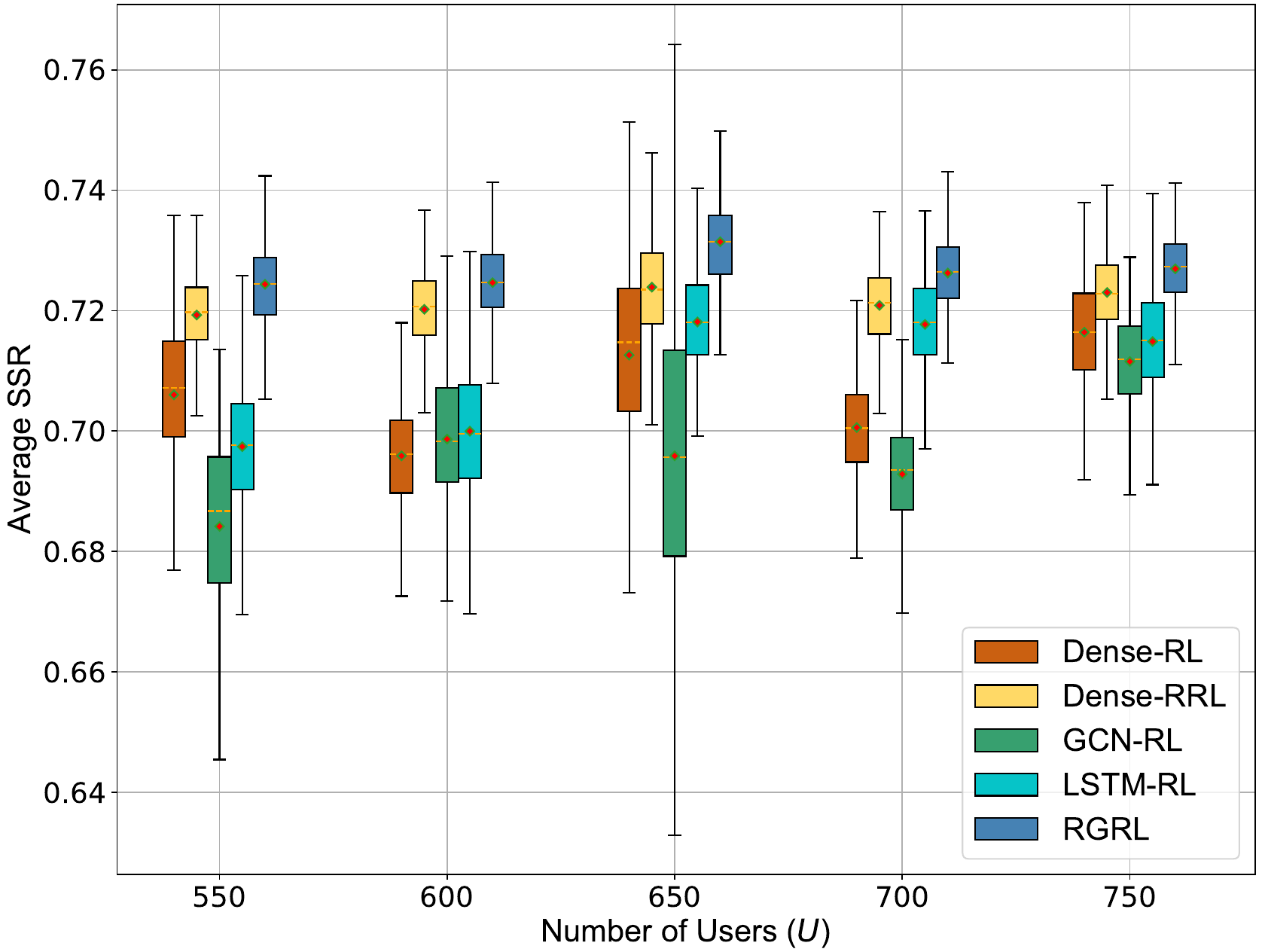}
\caption{Performance comparisons of the algorithms on different number of users under the use case 2. ($N=8$, $C_{\rm B}=10 \; {\rm GHz}$, $Z_{\rm B}=10$)}
\label{fig_6.12}
\end{figure}
\begin{figure}[!t]
\centering
\includegraphics[width=0.4\textwidth]{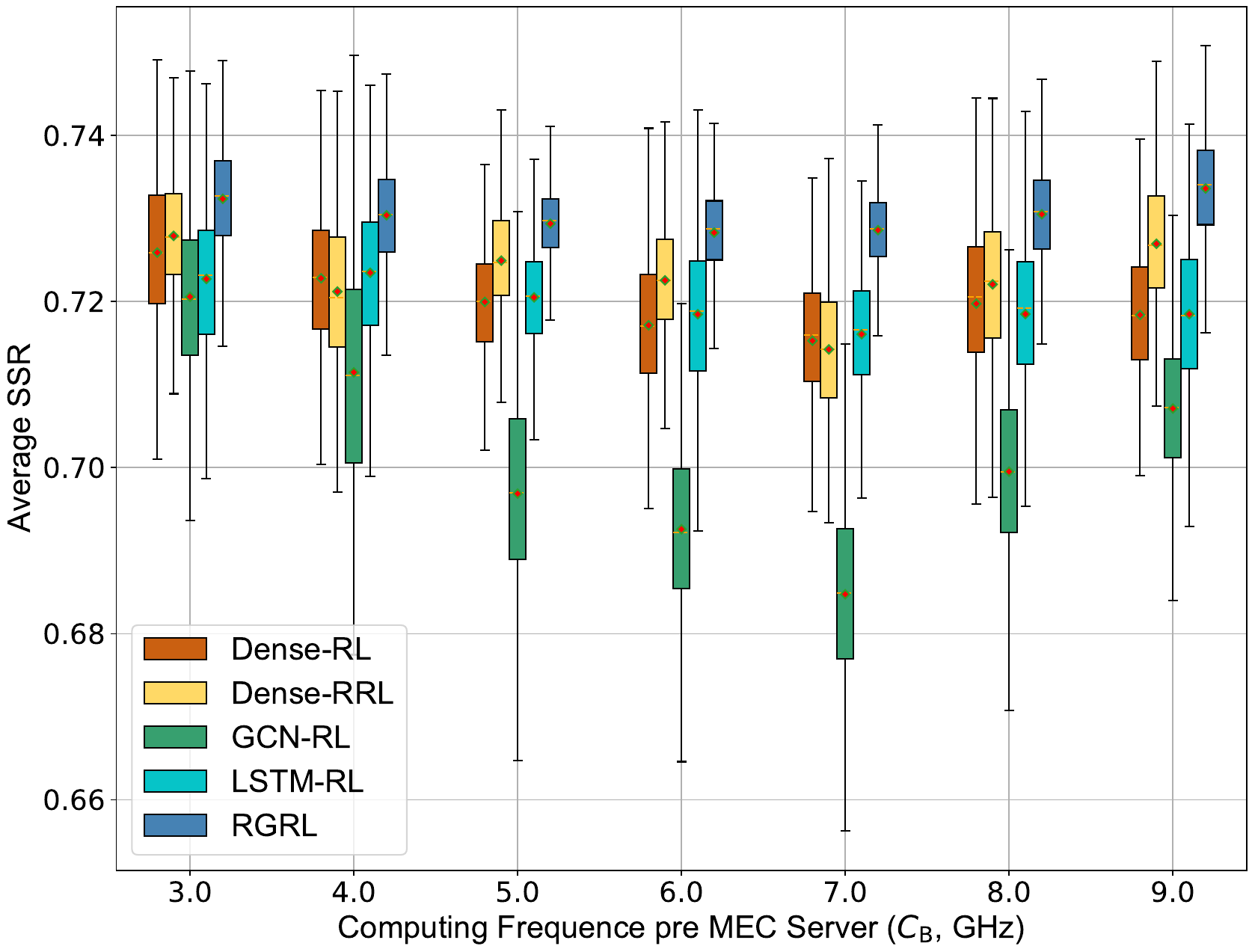}
\caption{Performance comparisons of the algorithms on different computing frequency under the use case 2. ($N=8$, $U=650$, $Z_{\rm B}=10$)}
\label{fig_6.13}
\end{figure}

\subsection{Simulation Results Under The Use Case 2}

In this subsection, we evaluate the proposed RGRL algorithm under the non-cooperative multi-node MEC-assisted RAN slicing scenario, where $A_{\rm max}=0$. Fig. \ref{fig_6.12} illustrates the performance comparisons of all the algorithms on different number of users under this use case. It can be seen from Fig. \ref{fig_6.12} that the proposed RGRL algorithm outperforms all the baseline algorithms in terms of average SSR, which further demonstrates the universal superiority of the proposed RGRL algorithm in different scenarios. We can also observe from Fig. \ref{fig_6.12} that the proposed RGRL algorithm is superior to the baseline Dense-RRL algorithm in terms of the average SSR. The reason is that the non-cooperative multi-node system considered in this use case can be abstracted as a null graph which still has a certain spatiality though the cooperations among nodes are disabled. The performance comparisons of all the algorithms at different computing frequencies $C_{\rm B}$ are provided in Fig. \ref{fig_6.13}. It is not difficult to observe from Fig. \ref{fig_6.13} that the proposed RGRL algorithm still outperform all the baseline algorithms regardless of the value of $C_{\rm B}$. In addition, by comparing the performance of the proposed RGRL algorithm, baseline Dense-RRL algorithm and baseline GCN-RL algorithm in Fig. \ref{fig_6.13}, we can infer that the GCN provides limited performance gain in the non-cooperative multi-node system, while the performance gain provided by the proposed time recurrent reinforcement learning framework is rather significant. This is because the temporal features are more significant than the spatial features in the non-cooperative multi-node MEC slicing system.

\begin{figure}[!t]
\centering
\includegraphics[width=0.4\textwidth]{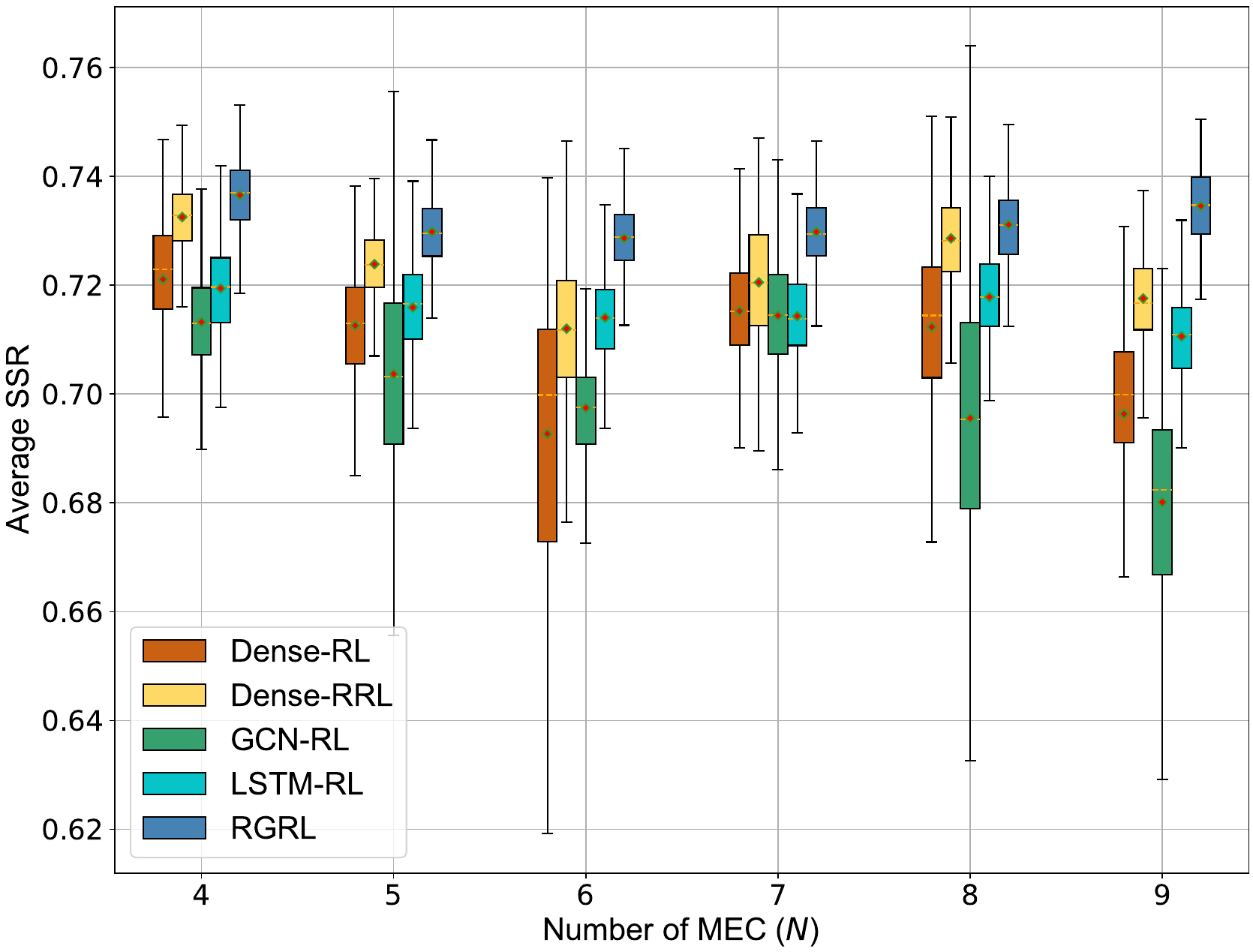}
\caption{Performance comparisons of the algorithms on different number of ENs under the use case 2. ($U=650$, $C_{\rm B}=10 \; {\rm GHz}$, $Z_{\rm B}=10$)}
\label{fig_6.14}
\end{figure}

\begin{figure}[!t]
\centering
\includegraphics[width=0.39\textwidth]{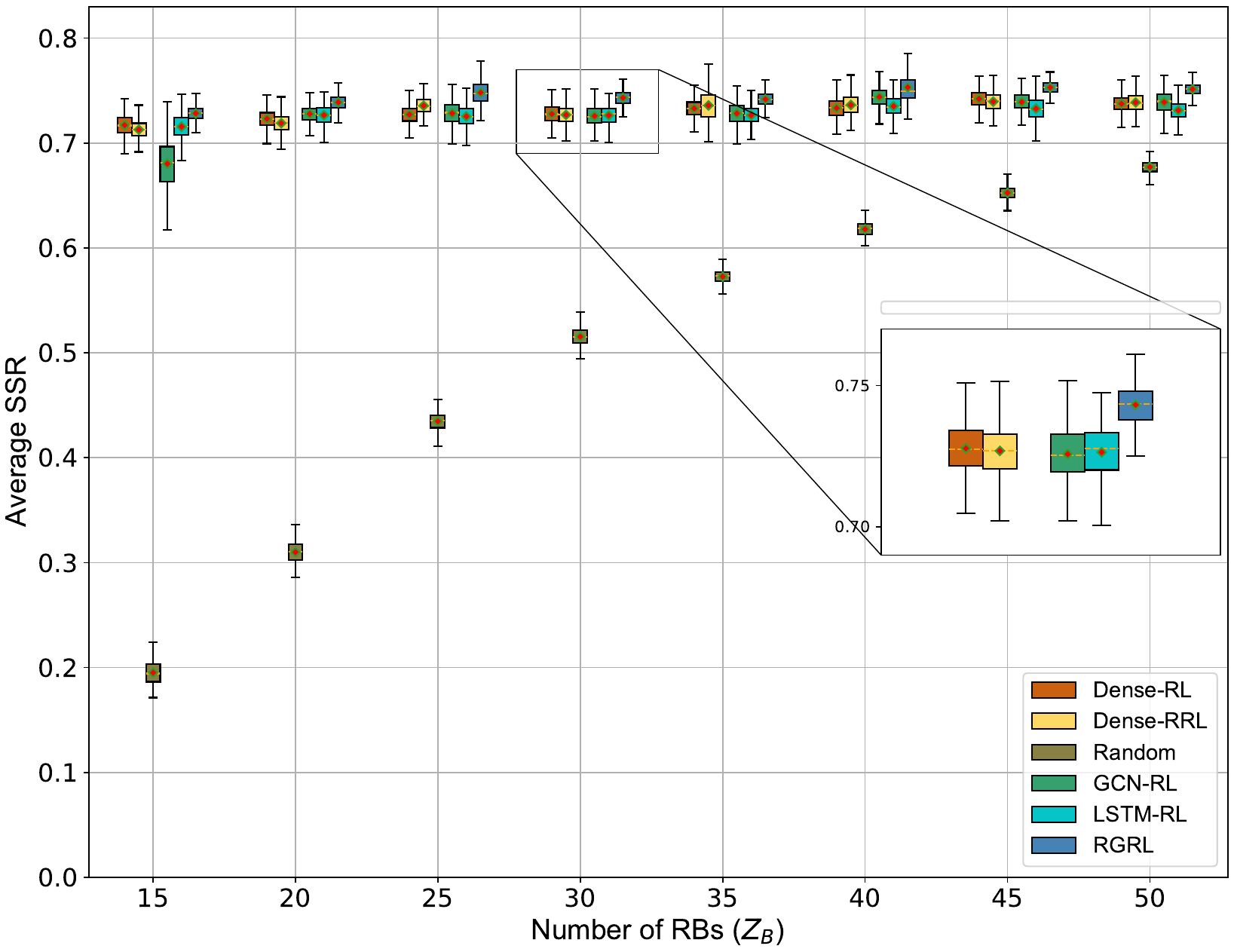}
\caption{Performance comparisons of the algorithms on different number of RBs under the use case 2. ($N=8$, $U=650$, $C_{\rm B}=10 \; {\rm GHz}$)}
\label{fig_6.15}
\end{figure}

Fig. \ref{fig_6.14} and Fig. \ref{fig_6.15} respectively provide the performance comparisons of all the algorithms on different $N$ and different $Z_{\rm B}$. It can be seen from these two figures that the proposed RGRL algorithm is superior to all the baseline algorithms in terms of the average SSR regardless of the value of $N$ and $Z_{\rm B}$. In general, through above simulation results in the three different scenarios, it can be confirmed that the proposed RGRL algorithm has universal advantages and can be used to improve the SSR performance when future wireless networks provide heterogeneous services.

\section{CONCLUSION}\label{sec_6}

In this paper, the policy optimization of hybrid RA is investigated in a collaborative multi-node MEC NS system to maximize the average SSR over a continous period of time. We abstracted the system into a weighted undirected graph and then proposed a RGRL algorithm to learn the optimal hybrid RA policy. Furthermore, we discuss the universality of the proposed RGRL algorithm by applying it to some classical use case scenario. Finally, The superior SSR performance, lower network complexity as well as the universality of the proposed RGRL algorithm are verified by numerical results.


\begin{thebibliography}{99}

\bibitem{WangZ22}
Z.\ Wang, Y.\ Wei, F.~R.\ Yu, and Z.\ Han, ``Utility optimization for resource allocation in multi-access edge network slicing: A twin-actor deep deterministic policy gradient approach,'' {\em IEEE Trans. Wireless Commun.}, vol.\ 21, no.\ 8, pp.\ 5842--5856, Aug.\ 2022.

\bibitem{Foukas17}
X.\ Foukas, G.\ Patounas, A.\ Elmokashfi, and M.~K.\ Marina, ``Network slicing in 5G: Survey and challenges,'' {\em IEEE Commun. Mag.}, vol. 55, no.\ 5, pp.\ 94--100, May 2017.

\bibitem{ZhouX16}
X.\ Zhou, R.\ Li, T.\ Chen, and H.\ Zhang, ``Network slicing as a service: Enabling enterprises\'own software-defined cellular networks,'' {\em IEEE Commun. Mag.}, vol.\ 54, no.\ 7, pp.\ 146--153, Jul.\ 2016.

\bibitem{YouX21}
X. You et al., ``Towards 6G wireless communication networks: Vision, enabling technologies, and new paradigm shifts,'' {\em Sci. China Inf. Sci.}, vol.\ 64, no.\ 1, pp.\ 1--74, Jan.\ 2021.


\bibitem{ITU-R17}
{\em Minimum Requirements Related to Technical Performance for IMT-2020 Radio Interface (s)}, document ITU-R M.2410--0, Nov.\ 2017.

\bibitem{ZaidiAA18}
A.~A.\ Zaidi, R.\ Baldemair, V.\ Moles-Cases, and et al., ``OFDM numerology design for 5G new radio to support IoT, eMBB, and MBSFN,'' {\em EEE Commun. Stand. Mag.}, vol.\ 2, no.\ 2, pp.\ 78--83, Jun.\ 2018.

\bibitem{KovalchukovR22}
R.\ Kovalchukov and et al., ``DECT-2020 new radio: The next step toward 5G massive machine-type bommunications,'' {\em IEEE Commun. Mag.}, vol.\ 60, no.\ 6, pp.\ 58--64, Jun.\ 2022.

\bibitem{AzariA19}
A.\ Azari, M.\ Ozger, and C.\ Cavdar, ``Risk-aware resource allocation for uRLLC: Challenges and strategies with machine learning,'' {\em IEEE Commun. Mag.}, vol.\ 57, no.\ 3, pp.\ 42--48, Mar.\ 2019.


\bibitem{TalebT17}
T.\ Taleb, K.\ Samdanis, B.\ Mada, and et al., ``On multi-access edge computing: A survey of the emerging 5G network edge cloud architecture and orchestration,'' {\em IEEE Commun. Surveys Tuts.}, vol.\ 19, no.\ 3, pp.\ 1657--1681, 3rd Quart., 2017.

\bibitem{ChienHT19}
H.-T.\ Chien, Y.-D.\ Lin, C.-L.\ Lai, and C.-T.\ Wang, ``End-to-end slicing as a service with computing and communication resource allocation for multi-tenant 5G systems,'' {\em IEEE Wireless Commun.} , vol.\ 26, no.\ 5, pp.\ 104--112, Oct.\ 2019.


\bibitem{FengJ20}
J.\ Feng, Q.\ Pei, F.~R.\ Yu, and et al., ``Dynamic network slicing and resource allocation in mobile edge computing systems,'' {\em IEEE Trans. Veh. Technol.}, vol.\ 69, no.\ 7, pp.\ 7863--7878, Jul.\ 2020.

\bibitem{ZhengC22}
C.\ Zheng, S.\ Liu, Y.\ Huang, and et al., ``Unsupervised recurrent federated learning for edge popularity prediction in privacy-preserving mobile-edge computing networks,'' {\em IEEE Internet Things J.}, vol.\ 9, no.\ 23, pp.\ 24328--24345, Dec.\ 2022.

\bibitem{Opti19}
X.\ Chen, H.\ Zhang, C.\ Wu, and et al., ``Optimized computation offloading performance in virtual edge computing systems via deep reinforcement learning,'' {\em IEEE Internet Things J.}, vol.\ 6, no.\ 3, pp.\ 4005--4018, Jun.\ 2019.

\bibitem{Combine22}
W.~K.~G.\ Seah, C.-H.\ Lee, Y.-D.\ Lin, and Y.-C.\ Lai, ``Combined communication and computing resource scheduling in sliced 5G multi-access edge computing systems,'' {\em IEEE Trans. Veh. Technol.}, vol.\ 71, no.\ 3, pp.\ 3144--3154, Mar.\ 2022.

\bibitem{Dyna20}
J.\ Feng, Q.\ Pei, F.~R.\ Yu, and et al., ``Dynamic network slicing and resource allocation in mobile edge computing systems,'' {\em IEEE Trans. Veh. Technol.}, vol.\ 69, no.\ 7, pp 7863--7878, Jul.\ 2020.

\bibitem{Utility22}
Z.\ Wang, Y.\ Wei, F.~R.\ Yu, and Z.\ Han, ``Utility optimization for resource allocation in multi-access edge network slicing: A twin-actor deep deterministic policy gradient approach," {\em IEEE Trans. Wireless Commun.}, vol.\ 21, no.\ 8, pp.\ 5842--5856, Aug.\ 2022.

\bibitem{Blockchain23}
T.\ Kwantwi, G.\ Sun, N.~A.~E.\ Kuadey, and et al., ``Blockchain-based computing resource trading in autonomous multi-access edge network slicing: A dueling double deep Q-learning approach," {\em IEEE Trans. Netw. Serv. Manage.}, to be published, doi: 10.1109/TNSM.2023.3240301.

\bibitem{Delay21}
S.\ Zarandi, and H.\ Tabassum, ``Delay minimization in sliced multi-cell mobile edge computing (MEC) systems,'' {\em IEEE Commun. Lett.}, vol.\ 25, no.\ 6, pp.\ 1964--1968, Jun.\ 2021.



\bibitem{WuZ21}
Z.\ Wu, S.\ Pan, F.\ Chen, and et al., ``A comprehensive survey on graph neural networks,'' {\em IEEE Trans. Neural Networks Learn. Syst.}, vol.\ 32, no.\ 1, pp.\ 4--24, Jan.\ 2021.

\bibitem{WangZ22}
Z.\ Wang, M.\ Eisen, and A.\ Ribeiro, ``Learning decentralized wireless resource allocations with graph neural networks,'' {\em IEEE Trans. Signal Process.}, vol.\ 70, pp.\ 1850--1863, Mar.\ 2022.


\bibitem{EisenM20}
M.\ Eisen, and A.\ Ribeiro, ``Optimal wireless resource allocation with random edge graph neural networks,'' {\em IEEE Trans. Signal Process.}, vol.\ 68, pp.\ 2977--2991, Apr.\ 2020.

\bibitem{WangD23}
D.\ Wang, Y.\ Bai, G.\ Huang, and et al., ``Cache-aided MEC for IoT: Resource allocation using deep graph reinforcement learning,'' {\em IEEE Internet Things J.}, vol.\ 10, no.\ 13, pp.\ 11486--11496, Jul.\ 2023.




\bibitem{Design20}
G.\ Faraci, C.\ Grasso, and G.\ Schembra, ``Design of a 5G network slice extension with MEC UAVs managed with reinforcement learning,'' {\em IEEE J. Sel. Areas Commun.}, vol.\ 38, no.\ 10, pp.\ 2356--2371, Oct.\ 2020.

\bibitem{HuaY20}
Y.\ Hua, R.\ Li, Z.\ Zhao, and et al., ``GAN-powered deep distributional reinforcement learning for resource management in network slicing,'' {\em IEEE J. Sel. Areas Commun.}, vol.\ 38, no.\ 2, pp.\ 334--349, Feb.\ 2020.

\bibitem{YanM19}
M.\ Yan, G.\ Feng, J.\ Zhou, and et al., ``Intelligent resource scheduling for 5G radio access network slicing,'' {\em IEEE Trans. Veh. Technol.}, vol.\ 68, no.\ 8, pp.\ 7691--7703, Aug.\ 2019.

\bibitem{CuiY22}
Y.\ Cui, X.\ Huang, P.\ He, and et al., ``QoS guaranteed network slicing orchestration for internet of vehicles,'' {\em IEEE Internet Things J.}, vol.\ 9, no.\ 16, pp.\ 15215--15227, Aug.\ 2022.

\bibitem{HaoM21}
M.\ Hao, D.\ Ye, S.\ Wang, and et al.,``URLLC resource slicing and scheduling in 5G vehicular edge computing,'' {\em in Proc. IEEE 93$^{\rm{rd}}$ Veh. Technol. Conf. (VTC2021-Spring)}, Helsinki, Finland, 2021, pp.\ 1--5.

\bibitem{GongY23}
Y.\ Gong, Y.\ Wei, F.~R.\ Yu, and Z.\ Han, ``Slicing-based resource optimization in multi-access edge network using ensemble learning aided DDPG algorithm,'' {\em J. Commun. Networks}, vol.\ 25, no.\ 1, pp.\ 1--14, Feb.\ 2023.

\bibitem{LiR20}
R.\ Li, C.\ Wang, Z.\ Zhao, and et al., ``The LSTM-based advantage actor-critic learning for resource management in network slicing with user mobility," {\em IEEE Commun. Lett.}, vol.\ 24, no.\ 9, pp.\ 2005--2009, Sept.\ 2020.



\bibitem{Distributed22}
S.\ Liu, C.\ Zheng, Y.\ Huang, and T.~Q.~S.\ Quek,``Distributed reinforcement learning for privacy-preserving dynamic edge caching,'' {\em IEEE J. Sel. Areas Commun.}, vol.\ 40, no.\ 3, pp.\ 749--760, Mar.\ 2022.

\bibitem{Hybrid21}
C.\ Zheng, S.\ Liu, Y.\ Huang, and L.\ Yang, ``Hybrid policy learning for energy-latency tradeoff in MEC-assisted VR video service,'' {\em IEEE Trans. Veh. Technol.}, vol.\ 70, no.\ 9, pp.\ 9006--9021, Sept.\ 2021.



\bibitem{JiangJ20}
J.\ Jiang, C.\ Dun, T.\ Huang, and Z.\ Lu, ``Graph convolutional reinforcement learning,'' {\em Proc. 8$^{\rm{th}}$ Int. Conf. Learn. Represent. (ICLR'20)}, Addis Ababa, Ethiopia, 2020, pp.\ 1665-1673.


\bibitem{YouY20}
Y.\ You, T.\ Chen, Z.\ Wang, and Y.\ Shen, ``L2-GCN: Layer-wise and learned efficient training of graph convolutional networks,'' {\em Proc. IEEE/CVF Conf. Comput. Vision and Pattern Recognit. (CVPR'20)}, Seattle, WA, USA, 2020, pp.\ 2124--2132.


\bibitem{Silver2014Deterministic}
D.\ Silver, G.\ Lever, N.\ Heess, T.\ Degris, D.\ Wierstra, and M.\ Riedmiller, ``Deterministic policy gradient algorithms,'' in {\em Proc. 31$^{\rm{st}}$ Int. Conf. Mach. Learn. (ICML'14)}, Beijing, China, June 2014, pp.\ 387--395.

\bibitem{Lillicrap2015Continuous}
T.~P.\ Lillicrap, J~J.\ Hunt, A.\ Pritzel, and et al., ``Continuous control with deep reinforcement learning,'' in {\em Proc. 4$^{\rm{th}}$ Int. Conf. Learn. Represent. (ICLR'16)}, San Juan, Puerto Rico, May 2016.




\bibitem{MeiJ21}
J.\ Mei, X.\ Wang, K.\ Zheng, and et al., ``Intelligent radio access network slicing for service provisioning in 6G: A hierarchical deep reinforcement learning approach,'' {\em IEEE Trans. Commun.}, vol.\ 69, no.\ 9, pp.\ 6063--6078, Sept.\ 2021.

\bibitem{Navarro20}
J.\ Navarro-Ortiz, and et al., ``A survey on 5G usage scenarios and traffic models,'' {\em IEEE Commun. Surv. Tutor.}, vol.\ 22, no.\ 2, pp.\ 905--929, Feb.\ 2020.

\bibitem{GSA2017}
GSA, ``5G network slicing for vertical industries,'' 2017.



\bibitem{SinhaS20}
S.\ Sinha, H.\ Bharadhwaj, A.\ Srinivas, and A,\ Garg, ``D2RL: Deep dense architectures in reinforcement learning,'' 2020, 	arXiv:2010.09163.

\bibitem{LiX15}
X.\ Li, L.\ Li, J.\ Gao, and et al., ``Recurrent reinforcement learning: A hybrid approach,'' 2015, arXiv:1509.03044.



\bibitem{Kingma2014Adam}
D.\ Kingma and J.\ Ba, ``Adam: A method for stochastic optimization,'' {\em Proc. 3$^{\rm{rd}}$ Int. Conf. Learn. Represent. (ICLR'15)}, San Diego, CA, USA, May 2015.



\end{thebibliography}
\end{document}